\documentclass[compact,tightcover,nostamp]{nuspaper}

\title{Agent Retrieval Bench: Evaluating Repository Context Retrieval for Coding Agents}
\shorttitle{Agent Retrieval Bench}
\date{July 2026}
\renewcommand{\printfrontmeta}{}

\author[1]{Bowen Qin}
\author[2]{Yi Xie}
\affil[1]{National University of Singapore (NUS)}
\affil[2]{Peking University}
\correspondence{qin.bowen@u.nus.edu}

\paperlinks{%
  \noindent\textbf{Project Page:}~\PaperLink{Project Website}{https://agent-retrieval-bench.github.io/}%
  \linksep\textbf{Artifacts:}~\PaperLink{Benchmark Releases}{https://huggingface.co/datasets/eyuansu71/agent_retrieval_bench}%
  \linksep\textbf{Source Code:}~\PaperLink{Evaluation Repository}{https://github.com/eyuansu62/agent-retrieval-bench}%
}

\newcommand{\arb}{Agent Retrieval Bench}
\newcommand{\code}[1]{\texttt{#1}}

\hypersetup{
  pdftitle={Agent Retrieval Bench: Evaluating Repository Context Retrieval for Coding Agents},
  pdfauthor={Bowen Qin and Xie Yi},
  pdfsubject={Repository context retrieval for coding agents},
  pdfkeywords={coding agents, repository retrieval, code search, context acquisition, benchmark}
}

\begin{document}

\maketitle

\begin{abstract}
Modern coding agents are usually evaluated by whether they eventually produce a correct patch, but patch generation depends on an earlier context-acquisition stage: the agent must find the repository files it needs to read. We introduce \arb{}, a file-level code retrieval benchmark for this upstream context-finding problem. Each sample is built from a real coding workflow signal and evaluated against a frozen base-commit repository; positive samples use the state before the resolving change. Unlike traditional code search, relevance in \arb{} is defined by what an agent needs next, not by direct semantic similarity between query and file. The benchmark covers four positive-retrieval tasks: \code{code2test}, \code{comment2context}, \code{trace2code}, and \code{edit2ripple}, which asks for additional files affected by an anchored change. A fifth subset tests selective retrieval with both natural evidence-backed no-gold cases and counterfactual wrong-repository controls.

The benchmark contains 427 samples across 25 repositories: 345 positive examples across the four retrieval tasks, 50 natural evidence-backed no-gold examples, and 32 counterfactual wrong-repository controls. Its combined corpus manifest contains 308 base-commit snapshots, 392K files, and 7.9M chunks; the released samples use 271 of those snapshots. We report lexical, vectorless RepoMap, open-source embedding, selective-abstention, and logged agent context-selection results. Validity diagnostics show complete corpus coverage, zero schema or corpus invalid rows, and zero fatal shortcut leakage from exact gold paths, raw patches, fix commits, or autogenerated review artifacts. On all 345 positive samples, Qwen3-Embedding-4B has the best sample-weighted MRR, Qwen3-Embedding-8B the best Recall@20, and RepoMap the best BCY@8k; per-task winners still differ sharply. Selective thresholds look useful when counterfactual controls are mixed with natural cases, but fail to improve selective success on the 50 natural no-gold cases, exposing a calibration gap. Logged trajectories over the 287 \code{code2test}, \code{comment2context}, and \code{trace2code} samples show that interactive agents still never touch any gold file on 27--35\% of samples. A 45-sample, single-run closed-tool seed-intervention pilot finds higher File F1 with less post-seed exploration for retrieval-derived initial context than for random non-gold context, while oracle gold context exposes substantial remaining headroom.
\end{abstract}

\keywords{coding agents; repository retrieval; code search; context acquisition; benchmark}

\section{Introduction}

Coding agents are often judged at the end of the workflow: did the generated patch pass tests, resolve the issue, or satisfy a benchmark oracle? This end-to-end measurement is useful, but it hides the context-acquisition layer that precedes editing. Before an agent can reason about a change, it must identify the files that matter. Modern agents may do this through a mixture of codebase indexes, repository maps, grep-like tools, file reads, and iterative search~\citep{aiderrepomap,yang2024sweagent}, but the underlying question is the same: which repository context should enter the model before it edits?

This makes agentic repository retrieval different from generic semantic code search. In traditional code search, relevance is usually defined by query-document semantic similarity, such as a natural-language description matching a function. In coding-agent workflows, relevance is defined by the next useful step in the software task. A pull request may describe an implementation change while the useful context is a regression test. A review comment may point at one file while the missing constraint lives in another module. A failure trace may mention a test file or stack frame while the root cause is an implementation file that does not look semantically similar to the error text. Agentic relevance is therefore often indirect, structural, or workflow-dependent.

Interactive exploration does not remove the need to evaluate this layer; it changes the cost model. An agent can compensate for a weak initial context by spending more tool calls, tokens, and latency on search. In our logged trajectory track over the 287 \code{code2test}, \code{comment2context}, and \code{trace2code} samples, an OpenAI strict-context agent reads 3.2 files per sample on average, while Codex CLI recovered contexts use about 6 file/path events per sample. Yet these trajectories still never touch any gold file on 35.2\% of OpenAI samples and 27.2--29.3\% of Codex samples. Conditional on samples where gold is eventually reached, the median first hit is step 2 for OpenAI and step 3 for Codex. These results show why static top-k scores should not be treated as full agent performance, while establishing context acquisition as a measurable upstream failure surface and cost center.

\arb{} isolates this upstream problem. \arb{} is a file-level benchmark: given a workflow-derived query and a repository corpus at the base commit, a retriever or context engine must rank files that an agent would need to read. The benchmark is deliberately not patch-level. It does not ask whether an agent can synthesize the final edit. Instead, it measures whether the context-acquisition layer can find useful files before editing begins, and whether it can do so under realistic rank and context-budget constraints.

Our claim is therefore component-level rather than end-to-end: retrieval quality changes the context acquired by a fixed retrieval policy and the cost of acquiring it. We do not claim that retrieval is the dominant cause of patch failure, that File F1 is a substitute for test-passing repair, or that the present results estimate the effect of retrieval on patch success. Establishing that link requires an executable repair benchmark with aligned outcome labels, which the current workflow-derived samples do not provide.

\begin{figure}[t]
\centering
\resizebox{\linewidth}{!}{%
\begin{tikzpicture}[
  font=\sffamily,
  stage/.style={
    draw=NUSRule!65,
    line width=0.55pt,
    rounded corners=2pt,
    fill=white
  },
  task/.style={
    stage,
    minimum width=5.0cm,
    minimum height=0.72cm,
    text width=4.65cm,
    align=left,
    inner xsep=0.16cm
  },
  target/.style={
    stage,
    minimum width=4.9cm,
    minimum height=0.72cm,
    text width=4.55cm,
    align=left,
    inner xsep=0.16cm
  },
  flow/.style={->, line width=0.8pt, draw=NUSBlue!80}
]
  \node[font=\bfseries\color{NUSBlue},anchor=west] at (0,6.45)
    {Workflow-derived query};
  \node[font=\bfseries\color{NUSBlue},align=center] at (7.55,6.45)
    {Frozen retrieval setting};
  \node[font=\bfseries\color{NUSBlue},anchor=west] at (10.25,6.45)
    {Agentic relevance target};

  \node[task,fill=NUSPaleBlue] (c2t) at (2.5,5.25)
    {\textbf{code2test}\quad PR intent or implementation change};
  \node[task,fill=NUSPaleOrange] (c2c) at (2.5,4.25)
    {\textbf{comment2context}\quad Review comment + given file};
  \node[task,fill=NUSPaleBlue] (t2c) at (2.5,3.25)
    {\textbf{trace2code}\quad Reproduced command + failure output};
  \node[task,fill=NUSPaleOrange] (e2r) at (2.5,2.25)
    {\textbf{edit2ripple}\quad Change intent + anchor diff/file};
  \node[task,fill=NUSPaleGray] (sel) at (2.5,1.25)
    {\textbf{selective}\quad Natural issue or wrong-repository control};

  \node[stage,fill=NUSPaleGray,minimum width=4.0cm,minimum height=4.72cm,
        text width=3.55cm,align=center] (setting) at (7.55,3.25) {%
    \textbf{\color{NUSDeepBlue}Repository snapshot}\\[0.10cm]
    \code{repo @ base\_commit}\\[-0.02cm]
    \footnotesize pre-resolution files only\\[0.24cm]
    \color{NUSRule!70}\rule{3.0cm}{0.4pt}\\[0.24cm]
    \normalsize\textbf{\color{NUSDeepBlue}Retriever / context\\engine}\\[0.08cm]
    \color{NUSInk}\footnotesize lexical \(\cdot\) embedding\\
    RepoMap \(\cdot\) agent tools\\[0.22cm]
    \color{NUSRule!70}\rule{3.0cm}{0.4pt}\\[0.20cm]
    \color{NUSInk}\footnotesize rank files under\\
    top-\(k\) or token budget};

  \node[target,fill=NUSPaleBlue] (o1) at (12.70,5.25)
    {\textbf{Tests}\quad related to the implementation change};
  \node[target,fill=NUSPaleOrange] (o2) at (12.70,4.25)
    {\textbf{Additional context}\quad excluding the given file};
  \node[target,fill=NUSPaleBlue] (o3) at (12.70,3.25)
    {\textbf{Root-cause source}\quad not merely visible test frames};
  \node[target,fill=NUSPaleOrange] (o4) at (12.70,2.25)
    {\textbf{Ripple files}\quad beyond the anchored edit};
  \node[target,fill=NUSPaleGray] (o5) at (12.70,1.25)
    {\textbf{Files or ABSTAIN}\quad when no local context is useful};

  \draw[flow] (c2t.east) -- (setting.west |- c2t.east);
  \draw[flow] (c2c.east) -- (setting.west |- c2c.east);
  \draw[flow] (t2c.east) -- (setting.west |- t2c.east);
  \draw[flow] (e2r.east) -- (setting.west |- e2r.east);
  \draw[flow] (sel.east) -- (setting.west |- sel.east);
  \draw[flow] (setting.east |- o1.west) -- (o1.west);
  \draw[flow] (setting.east |- o2.west) -- (o2.west);
  \draw[flow] (setting.east |- o3.west) -- (o3.west);
  \draw[flow] (setting.east |- o4.west) -- (o4.west);
  \draw[flow] (setting.east |- o5.west) -- (o5.west);

  \node[draw=NUSBlue,line width=0.7pt,fill=NUSAbstractBg,
        rounded corners=2pt,minimum width=15.15cm,minimum height=0.72cm,
        align=center,font=\bfseries\color{NUSDeepBlue}] at (7.58,-0.35)
    {Relevant = repository context the agent needs to read next,
     not the file most textually similar to the query};
\end{tikzpicture}%
}
\caption{\arb{} protocol and task-specific agentic relevance. Every method
receives a workflow-derived query and the repository frozen at its base commit,
then ranks files or abstains. Positive targets represent the context needed for
the next useful workflow step; selective cases test whether the system can
recognize that no repository-local context should be returned.}
\label{fig:arb-overview}
\end{figure}
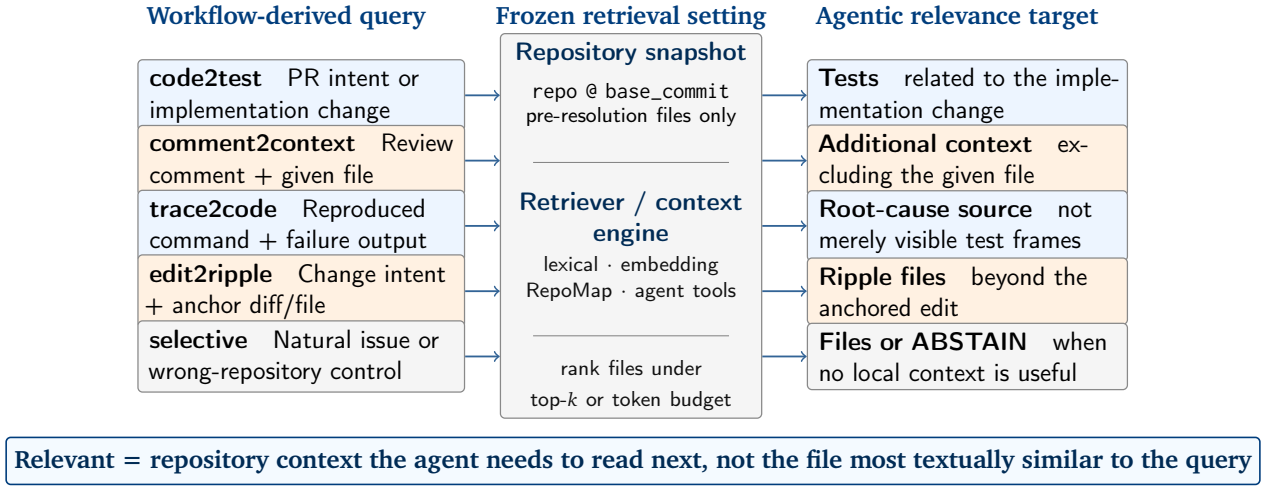

The benchmark contains four workflow-grounded positive-retrieval tasks. \code{code2test} asks for related tests from PR or implementation-change signals. \code{comment2context} asks for additional context files beyond the reviewed file. \code{trace2code} asks for root-cause source files from reproduced failure output, while treating visible tests as auxiliary context rather than main gold. \code{edit2ripple} provides an anchored change and asks for additional files affected by that change. A fifth selective subset combines 50 evidence-backed natural no-gold cases with 32 counterfactual wrong-repository controls. Each sample is evaluated against a base-commit corpus, preventing fixed-code leakage. Figure~\ref{fig:arb-overview} summarizes the shared protocol and the task-specific meaning of relevance.

This paper makes four contributions:
\begin{enumerate}
  \item We formalize \emph{agentic relevance} as the next repository context needed by a coding workflow, and operationalize it in four positive tasks plus natural and counterfactual no-gold retrieval.
  \item We construct and validate 427 samples against frozen base-commit corpora, with gold evidence, schema/corpus checks, and explicit controls for path, patch, commit, and generated-content leakage.
  \item We define an agent-oriented evaluation suite that combines ranked-file retrieval, token-budget BCY curves, selective abstention, trajectory cost, and a controlled-input context-seed pilot.
  \item We show empirically that no retrieval family dominates across workflow signals or context budgets: structural and semantic methods are complementary, Qwen3-4B and Qwen3-8B reverse order across \code{edit2ripple} budgets, and retrieval-derived seeds are more context-efficient than random context in the single-run intervention pilot.
\end{enumerate}

The goal is not to claim that bug localization, test traceability, or code review assistance are new in isolation. The contribution is a unified evaluation semantics and hygiene standard for the context-acquisition layer used by coding agents.

\section{Related Work}

\paragraph{Code search and code intelligence.}
CodeSearchNet~\citep{husain2019codesearchnet} and CodeXGLUE~\citep{lu2021codexglue} established influential benchmarks for semantic code search and broader code-intelligence tasks. These benchmarks evaluate representation learning and matching between natural language and code, often at the function or snippet level. \arb{} differs in both retrieval unit and query distribution: it evaluates file-level repository context retrieval from coding workflow signals, where the relevant file may be a test, a cross-module context file, or a root-cause source file that is not semantically close to the query.

\paragraph{Repository-level code context.}
RepoBench evaluates repository-level code auto-completion and highlights the importance of cross-file context~\citep{liu2023repobench}. \arb{} is complementary: instead of completing code at a known location, it asks which repository files an agent should inspect before editing. This shifts the problem from using context to finding context.

\paragraph{End-to-end coding-agent evaluation and localization.}
SWE-bench~\citep{jimenez2023swebench} and SWE-agent~\citep{yang2024sweagent} evaluate realistic software engineering agents on full issue-resolution workflows. Agentless~\citep{xia2024agentless} makes localization an explicit stage before repair, and LocAgent~\citep{chen2025locagent} studies graph-guided code localization for issue resolution. \arb{} targets a narrower but important intermediate stage. End-to-end patch success can fail because of retrieval, reasoning, editing, or validation. \arb{} isolates retrieval, making it possible to diagnose whether the agent found the necessary files before patch generation.

\paragraph{Agentic retrieval and exploration benchmarks.}
Recent work has begun to isolate context finding from patch generation. CORE-Bench evaluates code retrieval for agentic coding across code understanding, issue-to-edit localization, and broader context retrieval~\citep{zhang2026corebench}. SWE-Explore evaluates repository exploration under a fixed line budget, derives line-level ground truth from successful agent trajectories, and reports that its exploration metrics track downstream repair behavior~\citep{zhang2026sweexplore}. \arb{} is closest in spirit to this line, but occupies a different point in the design space. CORE-Bench emphasizes large-scale retrieval from issue and user-request style inputs, including edit-location and broader-context labels. SWE-Explore emphasizes adaptive line-level exploration, line-budget efficiency, and downstream association. \arb{} instead focuses on leakage-controlled \emph{workflow-signal} file retrieval: PR-to-test, review-comment-to-context, trace-to-root-cause, and anchored-edit-to-ripple queries, together with construction-separated no-gold controls. Its primary gold definition is the file an agent needs to read next, so visible stack frames and already-given files are not automatically successes. It trades scale and downstream repair validation for workflow-specific semantics, base-commit hygiene, fatal shortcut diagnostics, realistic distractors, context-budget curves, selective retrieval, and a controlled-input seed pilot that measures context acquisition only.

\begin{table}[t]
\centering
\caption{Positioning relative to nearby repository localization and agentic retrieval settings.}
\label{tab:nearby-work}
\small
\begin{tabularx}{\linewidth}{>{\raggedright\arraybackslash}p{0.18\linewidth}>{\raggedright\arraybackslash}p{0.19\linewidth}>{\raggedright\arraybackslash}p{0.23\linewidth}>{\raggedright\arraybackslash}X}
\toprule
Work & Query signal & Target relevance & Distinguishing evaluation focus \\
\midrule
SWE-bench / Agentless & Issue statements & Edit locations or files & Patch-pipeline localization \\
LocAgent & Issue statements & Code entities / locations & Graph-guided multi-hop localization \\
CORE-Bench & User and issue requests & Code, edit sites, context & Scale and broad retrieval coverage \\
SWE-Explore & Issue statements & Line regions from successful trajectories & Line-budget exploration and downstream repair association \\
\arb{} & PR, review, trace, and anchored-edit signals & Next-context files or abstention & Workflow semantics, base-commit hygiene, BCY/PES, selective retrieval \\
\bottomrule
\end{tabularx}
\end{table}

\paragraph{Traceability and test-code links.}
Software traceability studies links among requirements, code, tests, and other development artifacts~\citep{gotel2012traceability}. \code{code2test} is related to this tradition because the desired files are tests associated with an implementation or PR signal. The difference is that \arb{} does not reconstruct maintained trace matrices or coverage links. It asks whether a coding agent can infer which tests are useful next context from a workflow-derived query while avoiding direct test-path leakage.

\paragraph{Bug and fault localization.}
Bug-localization benchmarks and datasets such as Defects4J study mapping bug reports or failures to source locations~\citep{just2014defects4j}, and IR-based bug localization work studies source-file ranking from textual bug evidence~\citep{akbar2020irbuglocalization}. \arb{}'s \code{trace2code} task is adjacent, but it is framed around agent context retrieval from reproduced failure output. The gold files are root-cause source files needed for editing; visible tests and stack frames are evidence rather than automatically counted targets.

\paragraph{RAG and repository maps.}
Retrieval-augmented generation motivates retrieving evidence before generation~\citep{lewis2020rag}, and coding agents apply this idea to repository context. Practical systems such as Aider's repo map use structure and symbols rather than only vector similarity~\citep{aiderrepomap}. \arb{} provides an evaluation target for this setting and shows that semantic embeddings and structure-aware retrieval are complementary.

\section{Problem Definition}

An \arb{} sample consists of a repository, a base commit, a query, and a set of gold files. The candidate corpus is the repository at the base commit. A retriever receives the query and ranks candidate files. Metrics are computed at the file level; selective samples additionally specify why an empty gold set is expected. Appendix~\ref{app:sample-schema} gives the released evaluation-facing schema and one real sample from each positive task plus one natural no-gold sample.

This design follows the way coding agents consume context. A model typically needs readable source files, tests, configuration files, or related modules. Chunk-level retrieval is useful internally for scoring, but the benchmark's primary question is whether the correct file appears in the ranked file list.

\paragraph{Query.}
The query is derived from a real coding workflow signal: a PR title/body or implementation-change summary, a review comment plus the reviewed file, a reproduced failure command and failure excerpt, an anchored edit with change intent, or an issue whose resolution may lie outside the repository. Queries remove fatal shortcuts: final patches, raw fix diffs, fix commit hashes, model-generated suggestion or review-analysis artifacts, and exact gold paths. For trace-to-code retrieval, visible stack frames remain part of the realistic failure signal rather than automatic leakage.

\paragraph{Corpus.}
The corpus is built from the repository at the base commit. This is the state before the resolving fix. Candidate files are chunked for indexing and scoring, but ranked output is deduplicated to file paths.

\paragraph{Gold.}
Gold files are files an agent needs to read for the task. For \code{code2test}, gold files are related tests. For \code{comment2context}, gold files are additional context files; the reviewed file is recorded as given context and is not counted as main gold. For \code{trace2code}, gold files are root-cause source files; tests can be supporting evidence but are not main gold. For \code{edit2ripple}, gold files are additional affected source or test files beyond the anchor. Selective samples instead carry an empty gold set and a construction stratum; natural cases additionally carry an evidence-backed abstention reason.

\paragraph{Agentic relevance.}
We define a file as \emph{agentically relevant} when reading it would materially help the agent take the next correct step in the workflow, even if the query and file are not directly semantically similar. We distinguish four possible relations between the workflow signal and a gold file: \emph{semantic-direct} (\(D\)), where the query exposes target path or symbol cues; \emph{structural-indirect} (\(S\)), where an explicit repository relation connects a query anchor to the target; \emph{workflow-conventional} (\(W\)), where a workflow-role transformation such as implementation-to-test makes the target useful; and \emph{causal-indirect} (\(C\)), where a failure symptom or visible frame points to an evidence-backed root-cause implementation file. These relations are not mutually exclusive. We retain every detected relation and use a deterministic primary-label projection only to create disjoint analysis slices; Section~\ref{sec:relevance-relations} gives the predicates and precedence rule. The mechanism labels are generated by deterministic Codex-authored rules and are separate from the benchmark gold.

\paragraph{Metrics.}
The main metrics are Recall@5, Recall@10, Recall@20, MRR, and Budgeted Context Yield (BCY@B). Recall@k measures the fraction of gold files retrieved in the top k unique files. MRR uses the first ranked gold file. BCY@B asks how much labeled gold context fits inside a fixed token budget after a retriever ranks files. Our canonical packing protocol deduplicates ranked paths, renders each file from the released corpus as \code{\#\#\# \{path\}\textbackslash n\{content\}\textbackslash n}, greedily packs files by rank, and prefix-truncates only at the budget boundary. BCY uses a deterministic regex tokenizer, \code{regex\_code\_tokenizer\_v1}, which counts identifiers, number spans, and non-whitespace symbols as tokens. Path headers and separator newlines count against the budget. Formally, let \(G_i\) be the gold files for sample \(i\), \(L_{ig}\) the available corpus-text length of gold file \(g\), and \(m_{ig}(B)\) the number of its non-header content tokens packed under budget \(B\). For a minimum-content threshold \(\tau\),
\[
\mathrm{BCY}_{\tau}@B=\frac{1}{N}\sum_{i=1}^{N}\frac{1}{|G_i|}\sum_{g\in G_i}\mathbf{1}\!\left[m_{ig}(B)\geq\min(\tau,L_{ig})\right].
\]
Canonical BCY uses \(\tau=1\), measuring file exposure rather than full-file comprehension; the \(\min\) term gives complete short files credit. We report BCY curves at 4k, 8k, 16k, and 32k tokens, and use BCY@8k as the compact main-table point. Current details artifacts retain the top 20 unique files, so larger-budget points are computed from that stored prefix and are lower bounds whenever those files do not fill the budget. We treat non-gold context share as a diagnostic rather than a primary metric because useful but unjudged files may otherwise be counted as waste.

For logged trajectories, agent-native quantities are defined for a \emph{pair} of a retriever and an agent scaffold. Let \(A_i\) indicate whether the agent trajectory touches a gold file for sample \(i\), and let \(R_i@k\) indicate whether the retriever places any gold file in its top \(k\). We report the resulting 2-by-2 complementarity matrix: both hit, retriever-only, agent-only, and both miss. We also report Potential Exploration Savings, \(\mathrm{PES}@k = \frac{1}{N}\sum_i \max(T_i-1,0)\mathrm{I}[R_i@k]\), where \(T_i\) is the agent's first gold-touch event. PES is an upper-bound proxy for avoidable gold-localization delay, not a causal estimate of saved tool calls.

\section{Benchmark Tasks}

\subsection{\code{code2test}}

\code{code2test} simulates a coding agent receiving a PR or implementation-change signal and needing to identify relevant tests. The query may mention implementation files or behavior, but construction excludes exact test paths and final-patch evidence. The gold files are tests linked by released PR/change evidence.

\subsection{\code{comment2context}}

\code{comment2context} simulates a code review setting. The agent sees a review comment and the reviewed file. The benchmark asks for additional files needed to understand or satisfy the comment. The reviewed file is treated as given context, not as gold.

\subsection{\code{trace2code}}

\code{trace2code} simulates a reproduced failure. The query contains the command and failure excerpt. The gold files are root-cause source files. Tests and failure frames can be supporting context, but they are not counted as main gold unless they are supported as the root cause.

\subsection{\code{edit2ripple}}

\code{edit2ripple} simulates impact analysis after a localized edit is known. The query contains an anchor file, an anchor diff, and the change intent. The anchor file is given context rather than gold; the retriever must find additional source or test files affected by the change. Gold files are backed by changed-file evidence and recorded ripple relations such as shared symbols, component proximity, or required test updates.

\subsection{Selective No-Gold Retrieval}

The no-gold subset has two strata. Fifty \emph{natural} cases contain issue signals for which maintainer evidence attributes the resolution to an upstream dependency, external service, or user error. Thirty-two \emph{counterfactual wrong-repository} controls pair an issue signal with a repository that cannot contain its answer. Both strata have empty gold sets, but they test different phenomena and are therefore reported separately: the counterfactual controls test gross query--repository mismatch, whereas the natural cases test whether confidence can distinguish a plausible repository-local query from a genuinely external resolution.

\section{Dataset Construction}

The benchmark contains four positive retrieval tasks and two no-gold strata. Each sample is constructed from a repository workflow artifact, paired with task-specific relevance evidence, and evaluated against the repository state at a frozen base commit.

\subsection{Provenance and Construction Pipeline}
\label{sec:construction-pipeline}

We mine public GitHub development records from 25 purposively selected repositories spanning multiple languages and ecosystems. Repositories must have reconstructible Git history and usable pull-request, review, issue, or CI evidence; the collection is therefore a controlled benchmark sample rather than a random sample of GitHub. Crawls are bounded per repository and ordered by recently updated artifacts rather than restricted to one global calendar window. Source URLs, identifiers, and timestamps are retained when available.

Queries come from workflow evidence rather than model-written task descriptions. For \code{code2test}, we select merged pull requests that change both implementation and existing test files; the sanitized PR title/body and implementation-change summary form the query, while evidence-linked test files form the target. For \code{comment2context}, the query contains an inline review comment, its reviewed file, and sanitized hunk context. The target consists of additional base-existing files changed by the first valid post-comment response commit, excluding the reviewed file and requiring task-specific relation evidence. For \code{trace2code}, we reproduce a failing test at the base commit using test-side changes, then use the command and failure excerpt as the query; implementation files supported by the resolving change are root-cause targets, while visible tests remain auxiliary context. For \code{edit2ripple}, the anchor edit is given and other evidence-linked changed files are targets. Natural no-gold cases require maintainer evidence that the resolution lies outside the local repository; counterfactual controls pair a plausible signal with a deliberately wrong repository.

Construction then freezes the repository at the pre-resolution base commit and verifies that every positive target exists in that corpus. Query sanitization removes raw diffs, fix hashes, exact target paths, and autogenerated suggestion or review-analysis artifacts. A separate model-assisted evidence audit checks candidate labels against the stored workflow and resolving-change evidence; it does not generate query text and is independent of the retrieval methods evaluated in this paper. Schema, corpus-membership, duplication, and leakage checks run after assembly. These procedures establish traceable evidence for the benchmark labels, but do not imply that every potentially useful file has been exhaustively labeled.

\begin{table}[t]
\centering
\caption{Dataset composition. Positive tasks use file-level retrieval metrics; no-gold samples are evaluated by selective abstention.}
\label{tab:dataset-composition}
\begin{tabular}{llr}
\toprule
Subset & Evaluation role & Samples \\
\midrule
\code{code2test} & positive retrieval & 106 \\
\code{comment2context} & positive retrieval & 80 \\
\code{trace2code} & positive retrieval & 101 \\
\code{edit2ripple} & positive retrieval & 58 \\
\code{abstention-natural} & no-gold decision & 50 \\
\code{abstention-counterfactual} & wrong-repository control & 32 \\
\midrule
Positive subtotal & & 345 \\
Total & & 427 \\
\bottomrule
\end{tabular}
\end{table}

The release contains 427 samples across 25 repositories and uses 271 distinct repo/base pairs. The sample-bearing repositories span six primary languages: 13 Python, three Go, three Rust, three TypeScript, two Java, and one JavaScript repository. Appendix~\ref{app:repository-coverage} lists every repository and its per-task sample counts. The combined corpus manifest is a reusable superset with 308 repo/base rows across 29 repositories, 391,932 files, and 7,922,369 chunks.\footnote{The manifest preserves 37 reusable snapshots not referenced by the released samples; evaluation selects the 271 sample-used repo/base pairs.}

The corpus scale is intentionally much larger than the sample count. \arb{} evaluates whether a small set of evidence-backed workflow signals can retrieve files from large real repository snapshots. This makes the benchmark diagnostic rather than web-scale: each sample is expensive to validate, but each query searches a large, realistic candidate universe.

Exact text deduplication over the 287-sample \code{code2test}, \code{comment2context}, and \code{trace2code} evaluation corpora finds 1,118,431 unique embedding texts and 5,092,534 duplicate chunk texts, an 81.99\% duplicate fraction. This motivates shared corpus-embedding caches for practical evaluation because nearby base commits reuse many identical chunks.

\paragraph{Realistic distractors.}
\arb{} also preserves local distractors that are common in agent workflows. In \code{code2test}, queries may expose implementation files, symbols, or behavior while the target tests live elsewhere. In \code{comment2context}, the reviewed file is deliberately given to the agent but is not sufficient context and is not counted as main gold. In \code{trace2code}, visible stack frames and test paths remain in the failure signal, but the target is the root-cause source file. When available, construction records hard-negative files, and official evaluation uses the full file corpus rather than task-specific candidate filters. Thus a method must separate useful next context from nearby files that are visible, plausible, or semantically similar but insufficient.

\section{Quality and Leakage Controls}

\arb{}'s main risk is shortcut leakage. If a query contains the final patch, the fixing commit, or the exact gold path, the retrieval problem can collapse into string matching. We therefore separate released-data checks into three categories: gold-evidence traceability, corpus/schema validity, and shortcut diagnostics. These checks do not independently prove that the labeled set contains every useful file.

\paragraph{Gold evidence and corpus validity.}
All 345 positive samples carry released evidence linking their gold context to the workflow signal and base-commit corpus. The 287 \code{code2test}, \code{comment2context}, and \code{trace2code} samples record span/block-level evidence and query provenance; all 58 \code{edit2ripple} samples record changed-file ripple evidence. The selective subset contains 50 no-gold cases with maintainer-resolution evidence and 32 counterfactual wrong-repository controls. Automated release checks report zero invalid schema or corpus rows across the five subsets. This establishes evidence traceability and corpus membership, not exhaustive semantic correctness: alternative useful files may remain unlabeled.

\paragraph{Shortcut diagnostics.}
The fatal leakage diagnostic checks exact gold-path leakage, raw patch or diff leakage, fix-commit hash leakage, and model-generated suggestion or review-analysis artifacts. All fatal categories are zero. The last category is a sanitization check, not an indication that benchmark queries were generated by a model. We state the paper claim narrowly: the benchmark excludes direct answer leakage from exact paths, patches, commits, and generated review artifacts while preserving realistic workflow evidence.

\paragraph{Task diversity.}
The positive set contains 106 \code{code2test}, 80 \code{comment2context}, 101 \code{trace2code}, and 58 \code{edit2ripple} samples. We report task-level results because method rankings change by workflow signal and, for \code{edit2ripple}, by context budget. The 50 natural and 32 counterfactual no-gold samples are reported separately because their difficulty differs and abstention metrics are not commensurate with positive-only Recall or MRR.

\begin{table}[t]
\centering
\caption{Dataset validity diagnostics. Fatal leakage categories are exact shortcuts that would directly reveal a positive sample's answer.}
\label{tab:validity-diagnostics}
\begin{tabular}{lr}
\toprule
Diagnostic & Result \\
\midrule
Total samples & 427 \\
Evidence-backed positive gold & 345 / 345 \\
Natural evidence-backed no-gold decisions & 50 / 50 \\
Counterfactual wrong-repository controls & 32 / 32 \\
Schema or corpus invalid rows & 0 \\
Exact gold-path leakage & 0 \\
Raw patch / diff leakage & 0 \\
Fix-commit hash leakage & 0 \\
Autogenerated suggestion/review-artifact leakage & 0 \\
Positive task families with non-trivial method spread & 4 / 4 \\
\bottomrule
\end{tabular}
\end{table}

\section{Baselines}

We evaluate three static retrieval families. BM25 is a standard term-frequency lexical baseline with Robertson/Sparck Jones IDF and length normalization~\citep{robertson2009bm25}. The \code{lexical} row is a path-aware TF-IDF-style heuristic with path, basename, and symbol bonuses; it should not be read as standard BM25. RepoMap is a vectorless repository-structure baseline inspired by coding-agent repo maps. It ranks files using path, symbol, reference, and repository graph signals. Embedding baselines use open-source code/text embedding models: Qwen3-Embedding-4B and Qwen3-Embedding-8B~\citep{zhang2025qwen3embedding}, jina-code-embeddings-0.5b~\citep{kryvosheieva2025jina}, pplx-embed-v1-4b~\citep{eslami2026pplx}, and nomic-embed-code~\citep{suresh2025cornstack}. We use \emph{Qwen3-4B} and \emph{Qwen3-8B} as consistent display aliases for the two Qwen3 embedding checkpoints in main-text result tables, figures, and discussion; Appendix~\ref{app:embedding-implementation} retains their full repository identifiers. Hosted paid models are reported only in the separate logged context-selection track.

The embedding evaluator encodes queries and corpus chunks separately without an added query or passage prefix, L2-normalizes every vector, and ranks chunks by dot product, which is cosine similarity after normalization. It then deduplicates files in descending chunk-score order; equivalently, a file receives the maximum similarity of its chunks. Corpus inputs include the file path, chunk kind, optional symbol, and chunk content. These rows should therefore be read as prompt-free zero-shot baselines rather than model-card-optimized, task-prompted configurations. Appendix~\ref{app:embedding-implementation} gives checkpoint revisions, pooling, effective sequence limits, exact input rendering, and software versions.

All positive-task rows use \code{candidate\_filter=all\_files}. The main leaderboard covers all 345 positive samples and reports sample-weighted, four-task macro, and repository-macro averages. Repository-macro rank metrics first average samples within each repository and then average the 25 repositories equally. The relevance taxonomy, RRF, and trajectory joins cover the 287 \code{code2test}, \code{comment2context}, and \code{trace2code} samples for which the required labels and aligned artifacts are available. The 82 no-gold samples enter only the 427-sample selective-retrieval evaluation. We do not mix no-gold decisions into positive-only Recall or MRR.

\section{Main Results}

\subsection{Complete Positive Leaderboard}

Table~\ref{tab:overall-leaderboard} combines all four positive tasks. Weighted columns average over all 345 samples; macro columns give each task equal weight. Qwen3-4B has the best weighted MRR, Qwen3-8B has the best weighted Recall@20, and RepoMap has the best weighted BCY@8k. Qwen3-4B and Qwen3-8B retain the macro MRR and Recall@20 leads, respectively, while macro BCY is led by Qwen3-8B rather than RepoMap. Reporting both views therefore exposes, rather than hides, the effect of task frequency.

\begin{table*}[h!]
\centering
\caption{Complete positive leaderboard over 345 samples. Weighted metrics are sample-weighted over all four tasks; M-R@20, M-MRR, and M-BCY are unweighted task-macro averages. BCY uses canonical token-based file packing. The pplx row is provisional because of the tokenizer caveat in Appendix~\ref{app:embedding-implementation}.}
\label{tab:overall-leaderboard}
\scriptsize
\setlength{\tabcolsep}{5.0pt}
\begin{tabular}{lrrrrrrrr}
\toprule
Model & R@5 & R@10 & R@20 & MRR & BCY@8k & M-R@20 & M-MRR & M-BCY \\
\midrule
Qwen3-4B & 0.3182 & 0.4434 & 0.6306 & \textbf{0.2379} & 0.3409 & 0.6369 & \textbf{0.2441} & 0.3575 \\
Qwen3-8B & \textbf{0.3565} & \textbf{0.5428} & \textbf{0.7029} & 0.2336 & 0.3732 & \textbf{0.6988} & 0.2396 & \textbf{0.3892} \\
pplx-4B$^{\dagger}$ & 0.3066 & 0.4450 & 0.6072 & 0.2267 & 0.3549 & 0.6136 & 0.2347 & 0.3712 \\
RepoMap & 0.3134 & 0.4805 & 0.6333 & 0.2158 & \textbf{0.3788} & 0.6230 & 0.2150 & 0.3717 \\
Nomic & 0.2777 & 0.3741 & 0.5244 & 0.1986 & 0.2781 & 0.5266 & 0.2113 & 0.2954 \\
Jina-0.5B & 0.2449 & 0.3355 & 0.4823 & 0.1914 & 0.2783 & 0.5042 & 0.2024 & 0.2953 \\
Lexical & 0.2292 & 0.3548 & 0.4940 & 0.1574 & 0.2650 & 0.5072 & 0.1674 & 0.2840 \\
BM25 & 0.1899 & 0.3034 & 0.4452 & 0.1520 & 0.2051 & 0.4569 & 0.1551 & 0.2126 \\
\bottomrule
\end{tabular}
\end{table*}

The weighted MRR margin remains narrow: Qwen3-4B reaches 0.2379 and Qwen3-8B 0.2336; the provisional pplx run reaches 0.2267. Rank depth and context packing reverse parts of this ordering. Qwen3-8B reaches 0.7029 Recall@20, while RepoMap obtains the highest BCY@8k at 0.3788. This is why we retain per-task and context-budget analysis rather than collapsing the benchmark into one scalar.

The positive set is also repository-imbalanced: Gin contributes 88 of 345 samples, and the four largest repositories contribute 203 (58.8\%). Appendix Table~\ref{tab:repo-coverage} gives the complete repository-by-task distribution. Table~\ref{tab:repo-macro-sensitivity} therefore repeats the two rank metrics with repositories weighted equally. Qwen3-4B remains the MRR leader and becomes the Recall@20 leader under this view. Qwen3-8B's weighted Recall@20 advantage is thus partly concentrated in the more frequent repositories; it is not a repository-uniform win. This sensitivity check does not replace the sample-weighted task result, but prevents corpus composition from being hidden in a single overall number. We do not report repository-macro BCY because the canonical token-packed BCY artifacts are aggregated by task rather than stored per sample.

\begin{table}[h!]
\centering
\caption{Repository-imbalance sensitivity on 345 positive samples. W- columns weight samples equally; R- columns average within repository and then weight all 25 repositories equally.}
\label{tab:repo-macro-sensitivity}
\small
\begin{tabular}{lrrrr}
\toprule
Model & W-R@20 & R-R@20 & W-MRR & R-MRR \\
\midrule
Qwen3-4B & 0.6306 & \textbf{0.6344} & \textbf{0.2379} & \textbf{0.2501} \\
Qwen3-8B & \textbf{0.7029} & 0.6193 & 0.2336 & 0.2238 \\
pplx-4B$^{\dagger}$ & 0.6072 & 0.5714 & 0.2267 & 0.1871 \\
RepoMap & 0.6333 & 0.4619 & 0.2158 & 0.1298 \\
Nomic & 0.5244 & 0.4869 & 0.1986 & 0.1858 \\
Jina-0.5B & 0.4823 & 0.5424 & 0.1914 & 0.2031 \\
Lexical & 0.4940 & 0.3632 & 0.1574 & 0.1195 \\
BM25 & 0.4452 & 0.4318 & 0.1520 & 0.1336 \\
\bottomrule
\end{tabular}
\end{table}

\subsection{Task Winners}

The best model changes by task.

\par\medskip
\noindent\begin{minipage}{\textwidth}
\captionsetup{type=table,hypcap=false}
\captionof{table}{Positive-task winners by MRR and Recall@20. The aggregate row uses all 345 positive samples.}
\label{tab:task-winners}
\centering
\begin{tabular}{llrr}
\toprule
Task & Best MRR Model & MRR & Best R@20 \\
\midrule
Overall positive & Qwen3-4B & 0.2379 & Qwen3-8B \\
\code{code2test} & Qwen3-4B & 0.3225 & Qwen3-4B \\
\code{comment2context} & jina-code-embeddings-0.5b & 0.3043 & Qwen3-8B \\
\code{trace2code} & aider-style-repomap & 0.2742 & aider-style-repomap \\
\code{edit2ripple} & pplx-embed-v1-4b$^{\dagger}$ & 0.2877 & Qwen3-4B \\
\bottomrule
\end{tabular}
\end{minipage}
\par\medskip

These task winners support the central claim: different agentic retrieval signals require different inductive biases. A benchmark that reports only a single aggregate score would obscure this.

\subsection{Agentic Relevance Relations}
\label{sec:relevance-relations}

The relevance taxonomy makes the benchmark's central claim more concrete: agentic retrieval is not one relation between a query and a document. We operationalize the four relations as follows. Let \(Q_i\) be the set of non-path query tokens after identifier splitting and stop-token removal, and let \(H_i\) contain tokens from gold paths and released gold-symbol evidence. Directness is
\[
d_i=\frac{|Q_i\cap H_i|}{\max(1,\min(|Q_i|,|H_i|))}.
\]
Let \(s_i\) count recorded same-module, symbol/path-overlap, and source--test-path-overlap witnesses, plus same-module or stem links from a given/changed-file anchor to a gold path; let \(p_i\) indicate token overlap between an anchor path and a gold path. Let \(w_i\) count workflow witnesses such as same-PR changed tests or review-requested tests, a test-file gold, and the \code{code2test} workflow prior. The released annotations retain the exact witness-field names, summarized in Appendix Table~\ref{tab:sample-witness-fields} and instantiated by the examples that follow it.

Candidate relations may overlap. \(C\) is present for an evidence-backed \code{trace2code} symptom-to-root-cause pair; \(D\) and \(S\) remain possible secondary relations when their witnesses fire. For \code{code2test}, \(D\) is present when \(d_i\geq0.10\) or at least two query/gold cue tokens overlap, \(W\) is always present by task semantics, and \(S\) is present when \(s_i>0\) or \(p_i=1\). For \code{comment2context}, \(D\) is present when \(d_i\geq0.12\), at least two cue tokens overlap, or the released evidence records symbol/path overlap; \(W\) is present when \(w_i>0\); and \(S\) is present when \(s_i>0\), with \(S\) as the fallback when no other witness fires. The primary projection is deterministic: \code{trace2code} selects \(C\); \code{code2test} selects \(D\) only when \(d_i\geq0.30\), otherwise \(W\); and \code{comment2context} selects strong \(D\) when \(d_i\geq0.16\), otherwise \(W\) when \(w_i>s_i\), otherwise \(S\) when available, followed by the remaining detected relation. Thus \emph{primary} means dominant under a published rule, not exclusive membership or an ontological claim.

The taxonomy covers the 287 \code{code2test}, \code{comment2context}, and \code{trace2code} samples. Of these, 272 (94.8\%) have at least one secondary relation: 146 receive two detected relations, 126 receive three, and only 15 receive one. Table~\ref{tab:relevance-taxonomy} reports the primary projection so its rows form a partition. It contains direct semantic cases, workflow-conventional source-to-test and review-to-context cases, a smaller structural-indirect slice, and a causal-indirect trace slice. We leave \code{edit2ripple} outside this analysis because its anchor-to-ripple relation routinely combines structural and workflow evidence and has not been processed by these relation rules.

\begin{table}[h!]
\centering
\caption{Primary projection of the multi-label agentic-relevance taxonomy. Secondary relations are retained in the released annotations but omitted here so rows partition the 287 samples.}
\label{tab:relevance-taxonomy}
\begin{tabular}{lrrrr}
\toprule
Primary relation & Total & \code{code2test} & \code{comment2context} & \code{trace2code} \\
\midrule
Semantic-direct & 96 & 44 & 52 & 0 \\
Structural-indirect & 23 & 0 & 23 & 0 \\
Workflow-conventional & 67 & 62 & 5 & 0 \\
Causal-indirect & 101 & 0 & 0 & 101 \\
\bottomrule
\end{tabular}
\end{table}

The primary projection provides a descriptive view of family-level behavior behind the aggregate leaderboard. Table~\ref{tab:relevance-type-results} reports the best MRR and best Recall@20 method within each primary slice. For a file \(f\), reciprocal-rank fusion~\citep{cormack2009rrf} assigns \(s_{\mathrm{RRF}}(f)=\sum_{m\in\mathcal{M}}(60+r_m(f))^{-1}\), where \(r_m(f)\) is its one-based rank under method \(m\); a method contributes zero when \(f\) is absent from its stored top-20 list. We use \emph{RRF-3} for the three-way fusion of Qwen3-4B, Qwen3-8B, and RepoMap, and \emph{RRF-Q8+RM} for the two-way Qwen3-8B--RepoMap fusion. Semantic-direct and workflow-conventional primary cases favor embedding-heavy or fused rankers. Causal-indirect cases reverse the picture: RepoMap has the best MRR and RRF-Q8+RM has the best Recall@20, while lexical and grep baselines are also competitive. The structural-indirect slice is smaller and should be read as diagnostic rather than definitive. Because the projection is task-conditioned and the underlying relations overlap, these rows describe benchmark mechanisms; they do not identify a task-independent causal effect of a relevance type.

\begin{table}[h!]
\centering
\caption{Best methods by primary agentic-relevance projection. R@20 and MRR are computed within each disjoint projected slice.}
\label{tab:relevance-type-results}
\begin{tabular}{lrrr}
\toprule
Primary relation & n & Best MRR & Best R@20 \\
\midrule
Semantic-direct & 96 & 0.3939 RRF-3 & 0.7448 RRF-3 \\
Structural-indirect & 23 & 0.2213 Qwen3-4B & 0.5725 RRF-3/Q8+RM \\
Workflow-conventional & 67 & 0.2876 Qwen3-4B & 0.6463 RRF-3 \\
Causal-indirect & 101 & 0.2742 RepoMap & 0.8795 RRF-Q8+RM \\
\bottomrule
\end{tabular}
\end{table}

\subsection{Simple Hybrid Rank Fusion}

The three-task split suggests that semantic and structure-aware retrievers are complementary. To test whether this complementarity is visible without training a new model, we apply reciprocal-rank fusion (RRF, \(k=60\)) to the stored top-20 file rankings from Qwen3-8B and RepoMap. Because this post-hoc fusion only uses file rankings, we report rank metrics in the dedicated fusion table; its top-20-limited BCY is shown separately in the context-budget comparison. The table's nDCG@20 uses binary file relevance and the standard discounted cumulative-gain normalization~\citep{jarvelin2002ndcg}.

\begin{table}[t]
\centering
\caption{Post-hoc RRF fusion of Qwen3-8B and RepoMap on the 287 \code{code2test}, \code{comment2context}, and \code{trace2code} samples.}
\label{tab:rrf-fusion}
\begin{tabular}{lrrrr}
\toprule
Task & R@20 & MRR & Any@20 & nDCG@20 \\
\midrule
overall & 0.7331 & 0.2713 & 0.8153 & 0.3651 \\
\code{code2test} & 0.6972 & 0.2811 & 0.7642 & 0.3662 \\
\code{comment2context} & 0.5958 & 0.2635 & 0.7375 & 0.3136 \\
\code{trace2code} & 0.8795 & 0.2672 & 0.9307 & 0.4048 \\
\bottomrule
\end{tabular}
\end{table}

This simple fusion improves overall MRR from the best single-run value of 0.2296 to 0.2713 and improves overall Recall@20 from 0.7070 to 0.7331. On \code{trace2code}, it also raises Recall@20 above both inputs (0.8795 versus 0.8366 for RepoMap and 0.7970 for Qwen3-8B). The result does not make a single universal retriever: \code{code2test} MRR remains slightly below Qwen3-4B. Instead, it demonstrates that the benchmark exposes usable complementarity between semantic and structural signals.

\subsection{Context-Budget and Agent Context Selection}

Top-k retrieval metrics do not fully determine whether an agent can use the retrieved files under a realistic context budget. We therefore also report canonical BCY@8k and a matched top-3/final-context file-F1 comparison between static retrievers and logged agent trajectories.

\begin{table}[t]
\centering
\caption{Context-budget and context-selection results on \code{code2test}, \code{comment2context}, and \code{trace2code}. Static retrievers are measured by BCY@8k and matched top-3 file F1 where available; agent rows use logged final-context trajectories.}
\label{tab:context-selection}
\begin{tabular}{llcc}
\toprule
Method & Type & BCY@8k & Top-3 / final File F1 \\
\midrule
RRF(Qwen3-8B+RepoMap) & retriever & 0.4884 & -- \\
RepoMap & retriever & 0.3828 & 0.1102 \\
Qwen3-4B & retriever & 0.3079 & 0.1247 \\
Lexical & retriever & 0.2282 & 0.0721 \\
Grep strict & retriever & 0.2157 & 0.0841 \\
OpenAI GPT-5.4-mini strict & agent & -- & 0.3113 \\
Codex CLI GPT-5.4 & agent & -- & 0.3342 \\
Codex CLI GPT-5.5 & agent & -- & 0.3255 \\
\bottomrule
\end{tabular}
\end{table}

We test whether canonical BCY is artificially inflated by the one-token exposure convention by recomputing BCY@8k with \(\tau\in\{1,16,32,64,128\}\). Across all 12 methods evaluated on \code{code2test}, \code{comment2context}, and \code{trace2code}, the ordering is unchanged and the largest absolute decrease from \(\tau=1\) to 128 is 0.0081. The eight \code{edit2ripple} methods are likewise unchanged in order, with a maximum decrease of 0.0058. Thus the reported comparisons do not depend on crediting a boundary-truncated gold file after only one content token.

\subsection{File-versus-Span Granularity Sensitivity}

BCY measures whether a ranked file becomes visible under a context budget; it does not claim that the retriever has isolated the useful function or region within that file. This distinction matters in our data. Table~\ref{tab:gold-file-span-size} measures every sample--gold-file occurrence at its base commit. Across the 345 positive samples, the median gold file has 515 lines, 51.8\% have more than 500 lines, 24.1\% have more than 1,000 lines, and the maximum has 10,218 lines. All 287 samples from \code{code2test}, \code{comment2context}, and \code{trace2code} additionally have line-span annotations. Their median labeled evidence occupies only 27 lines, or 4.7\% of its containing file; 75.7\% of span-file pairs use at most 10\% of the file. Therefore, a file-level hit is useful evidence of document reachability, but can substantially overstate within-file localization precision.

\begin{table*}[t]
\centering
\caption{Gold-file size and annotated evidence density. Gold-file counts are sample--file occurrences in the base-commit corpus. Span statistics cover the 287 \code{code2test}, \code{comment2context}, and \code{trace2code} samples; \code{edit2ripple} does not yet have span annotations.}
\label{tab:gold-file-span-size}
\scriptsize
\setlength{\tabcolsep}{2.5pt}
\begin{tabular}{lrrrrrrr}
\toprule
Task & Gold files & Med. file & P90 file & Span pairs & Med. evidence & Med. evidence/file & $\leq$10\% file \\
\midrule
\code{code2test} & 148 & 518 & 1,506 & 139 & 14 lines & 4.1\% & 79.1\% \\
\code{comment2context} & 140 & 491 & 1,910 & 128 & 32 lines & 6.3\% & 70.3\% \\
\code{trace2code} & 127 & 725 & 3,020 & 124 & 27 lines & 3.9\% & 77.4\% \\
\code{edit2ripple} & 99 & 414 & 1,500 & -- & -- & -- & -- \\
\midrule
Overall & 514 & 515 & 2,011 & 391 & 27 lines & 4.7\% & 75.7\% \\
\bottomrule
\end{tabular}
\end{table*}

We also compare file-level retrieval with overlap against the annotated lines in Table~\ref{tab:file-span-sensitivity}. This diagnostic reuses the stored chunk rankings and their original 8,000-character packing rule, then computes line recall, precision, and F1 from the union of packed chunk intervals. It is intentionally labeled as a legacy sensitivity analysis rather than canonical token-based BCY. The method ordering changes: Qwen3-8B leads file Recall@20 at 0.7070 but ranks fourth by line F1 at 0.0165, while Qwen3-4B leads line F1 at 0.0276. RepoMap emits only file rankings and therefore has no line-resolved score. The absolute line precision and F1 values are low for every method. Thus our conclusion that retrieval families have complementary strengths survives, but the file-level leaderboard must not be interpreted as a solved span-localization problem.

\begin{table}[t]
\centering
\caption{File exposure versus line overlap on the 287-sample span-annotated core. ``Exposure'' and line metrics use the legacy 8,000-character chunk packing and are not canonical token BCY.}
\label{tab:file-span-sensitivity}
\scriptsize
\setlength{\tabcolsep}{3.5pt}
\begin{tabular}{lrrrrr}
\toprule
Method & File R@20 & Exposure & Line R & Line P & Line F1 \\
\midrule
Qwen3-4B & 0.6143 & 0.2516 & 0.0918 & 0.0216 & \textbf{0.0276} \\
pplx-embed-v1-4b & 0.5929 & 0.1359 & 0.0803 & 0.0186 & 0.0234 \\
jina-code-embeddings-0.5b & 0.4491 & 0.1568 & 0.0630 & 0.0136 & 0.0181 \\
Qwen3-8B & \textbf{0.7070} & 0.1533 & 0.0573 & 0.0137 & 0.0165 \\
nomic-embed-code & 0.5145 & 0.0832 & 0.0525 & 0.0110 & 0.0155 \\
BM25 & 0.4344 & 0.0993 & 0.0400 & 0.0086 & 0.0121 \\
Lexical & 0.4750 & 0.0738 & 0.0223 & 0.0044 & 0.0062 \\
RepoMap & 0.6367 & 0.0592 & -- & -- & -- \\
\bottomrule
\end{tabular}
\end{table}

\subsection{Trajectory Cost and Seed Intervention}

Static reachability and interactive context acquisition are complementary evaluation surfaces. An interactive agent can buy better final context by spending additional search and read events, but may still miss gold. We therefore summarize context-acquisition cost in the logged trajectories, join those trajectories to static rankings, and then intervene on initial context under a fixed closed-tool policy.

\begin{center}
\centering
\captionsetup{hypcap=false}
\captionof{table}{Logged context-acquisition cost on the 287 \code{code2test}, \code{comment2context}, and \code{trace2code} samples. Any-gold is the fraction of samples where any gold file appears in the logged context events. Median first hit is computed over successful samples. Codex rows use recovered file/path events, so they are cost proxies rather than wall-clock measurements.}
\label{tab:trajectory-cost}
\begin{tabular}{lrrr}
\toprule
Method & Events/sample & Any-gold & Median first hit \\
\midrule
OpenAI GPT-5.4-mini strict & 3.2 & 0.648 & 2 \\
Codex CLI GPT-5.4 & 6.2 & 0.728 & 3 \\
Codex CLI GPT-5.5 & 6.5 & 0.707 & 3 \\
\bottomrule
\end{tabular}
\end{center}

Read in reverse, Any-gold is also a miss-rate diagnostic. Even with interactive exploration, the logged agents never touch any gold file on 35.2\% of OpenAI strict-context samples and 27.2--29.3\% of Codex samples. The median first-hit numbers are therefore conditional on successful context acquisition, not evidence that all agents quickly find the needed files.

We then join the logged trajectories with static retriever ranks on the same sample IDs. The most informative subset is \emph{late-hit}: samples where the agent eventually reaches gold, but only after more than three context events. These are cases where a better starting context can plausibly reduce gold-localization delay.

\begin{center}
\centering
\captionsetup{hypcap=false}
\captionof{table}{Retriever coverage on late-hit samples from the three trajectory tasks. Late-hit samples are those where the logged agent first reaches a gold file after more than three context events.}
\label{tab:late-hit-join}
\begin{tabular}{llrrrr}
\toprule
Agent & Retriever & Samples & Events/sample & Any@10 & Any@20 \\
\midrule
Codex GPT-5.4 & Qwen3-8B & 77 & 7.6 & 0.675 & 0.818 \\
Codex GPT-5.4 & Qwen3-4B & 77 & 7.6 & 0.623 & 0.779 \\
Codex GPT-5.5 & Qwen3-8B & 73 & 8.1 & 0.726 & 0.863 \\
Codex GPT-5.5 & Qwen3-4B & 73 & 8.1 & 0.658 & 0.836 \\
\bottomrule
\end{tabular}
\end{center}

The same join also exposes complementarity rather than a one-sided ``retriever rescues agent'' story. Table~\ref{tab:complementarity} reports the 2-by-2 matrix for Codex CLI GPT-5.4 trajectories and top-20 retriever hits. Retriever-only cases are samples where static retrieval finds gold that the agent never touches; agent-only cases are the reverse, where interaction reaches gold that the static retriever misses. PES@20 is the potential localization-delay upper bound defined above.

\begin{center}
\centering
\captionsetup{hypcap=false}
\captionof{table}{Trajectory-conditioned complementarity for Codex CLI GPT-5.4 at top 20. Rows sum to one over the 287 samples from the three trajectory tasks.}
\label{tab:complementarity}
\begin{tabular}{lrrrrr}
\toprule
Retriever & Both hit & Retriever-only & Agent-only & Both miss & PES@20 \\
\midrule
Qwen3-4B & 0.509 & 0.178 & 0.220 & 0.094 & 1.213 \\
Qwen3-8B & 0.627 & 0.167 & 0.101 & 0.105 & 1.345 \\
RepoMap & 0.592 & 0.125 & 0.136 & 0.146 & 1.233 \\
RRF(Qwen3-8B+RepoMap) & 0.652 & 0.164 & 0.077 & 0.108 & 1.373 \\
\bottomrule
\end{tabular}
\end{center}

This cost view turns the agent-vs-retriever comparison into a sharper motivation for \arb{}. Qwen3-4B reaches 0.686 Any@20 as a static ranking, close to the OpenAI trajectory's 0.648 any-gold-read rate, but its matched top-3 context F1 is only 0.1247. On late-hit Codex samples, Qwen3-8B already places gold in the top 20 for more than 80\% of cases. The complementarity matrix shows the other side as well: interactive exploration still reaches gold in samples where a static retriever misses. The gap means that compact context selection, rank depth, and exploration cost are different evaluation surfaces. A better retrieval layer can seed the agent closer to useful files, while interaction remains necessary for cases static retrieval misses.

We further run a controlled-input seed intervention over 45 samples, 15 each from \code{code2test}, \code{comment2context}, and \code{trace2code}. The agent policy and tool budget are fixed: Docker-isolated Codex GPT-5.5 can only request evaluator-mediated \code{list\_dir}, \code{grep}, \code{read\_file}, and \code{submit} actions, with at most 16 tool calls, 20 model turns, 8k post-seed read tokens, 1,200 tokens per file, and a final set of three files. The only intended input intervention is the initial context seed. We compare no seed, a random non-gold seed, lexical and embedding seeds, an RRF hybrid seed, and an oracle seed containing gold files. The random arm controls for the effect of merely adding context; the oracle arm estimates headroom when localization is solved. Each sample--arm pair has one completed trajectory. The Codex CLI invocation fixes the model name but exposes neither an explicit temperature nor a random seed in the saved run configuration, so this is a descriptive pilot rather than a repeated-sampling estimate.

\begin{table*}[t]
\centering
\caption{Closed-tool seed-intervention pilot on 45 samples stratified across \code{code2test}, \code{comment2context}, and \code{trace2code}. All rows use the same Codex GPT-5.5 closed-tool policy; only the initial context seed changes. Post-seed tokens exclude preloaded seed context. One trajectory is observed per sample--arm pair.}
\label{tab:seed-intervention}
\small
\begin{tabular}{lrrrrrrrr}
\toprule
Arm & n & Final F1 & \(\Delta\)F1 & Any-gold & First hit & Calls & Post toks & Seed toks \\
\midrule
No seed & 45 & 0.3222 & 0.0000 & 0.5111 & 3.0 & 3.71 & 2137.3 & 0.0 \\
Random & 45 & 0.3437 & 0.0215 & 0.7333 & 3.0 & 8.49 & 3681.7 & 2427.8 \\
Lexical & 45 & 0.3981 & 0.0759 & 0.8000 & 2.0 & 6.24 & 2243.9 & 3061.1 \\
Qwen8B & 45 & 0.3726 & 0.0504 & 0.7556 & 1.0 & 5.51 & 2047.7 & 3204.2 \\
RRF & 45 & 0.3967 & 0.0744 & 0.8000 & 1.0 & 5.42 & 1856.7 & 3298.8 \\
Oracle & 45 & 0.6337 & 0.3115 & 1.0000 & 0.0 & 4.18 & 1736.4 & 1781.3 \\
\bottomrule
\end{tabular}
\end{table*}

Table~\ref{tab:seed-intervention} shows that extra context and prompt perturbation already help: random non-gold context improves final File F1 by 0.0215 and raises any-gold acquisition. This makes the intervention a conservative test of seed quality rather than a pure comparison against an inert control. The retrieval-derived seeds are stronger because they improve more efficiently than random context: lexical and RRF seeds add about 0.075 File F1, reach gold earlier, and require substantially fewer post-seed read tokens and tool calls than the random arm. Qwen3-8B reaches gold earliest but has a smaller final-F1 gain, showing that early gold contact and final context selection are related but distinct. Oracle gold context raises final File F1 by 0.3115, rescues every no-seed miss, and still does not reach perfect final precision.

\begin{table}[t]
\centering
\caption{Paired seed-intervention deltas against the no-seed arm. Rescue/no-seed miss is the fraction of no-seed misses that become any-gold hits under the seed.}
\label{tab:seed-intervention-paired}
\small
\begin{tabular}{lrrrr}
\toprule
Arm & Improved & Worsened & Rescue/miss & Seed-hit \(\Delta\)F1 \\
\midrule
Random non-gold & 0.2667 & 0.3333 & 0.5455 & 0.0000 \\
Lexical & 0.3333 & 0.2889 & 0.6364 & 0.0625 \\
Qwen3-8B & 0.2889 & 0.3556 & 0.5909 & 0.2119 \\
RRF(Qwen3-8B+RepoMap) & 0.4222 & 0.3333 & 0.7273 & 0.1154 \\
Oracle gold & 0.6444 & 0.1778 & 1.0000 & 0.3115 \\
\bottomrule
\end{tabular}
\end{table}

The paired deltas in Table~\ref{tab:seed-intervention-paired} are deliberately more conservative than aggregate averages: seeded runs can worsen some samples. The result is therefore not that retrieval always improves an agent. In these single-run trajectories under a fixed closed-tool policy, retrieval-derived seeds are associated with higher File F1 and earlier gold contact with less post-seed exploration than a random non-gold seed, while oracle context gives an upper bound on remaining localization headroom. Because policy sampling is not repeated, small differences between retrieval arms should not be interpreted as statistically resolved rankings.

\paragraph{Evidence boundary.}
The seed intervention is a functional validation of \emph{context acquisition}, not of software repair. The closed-tool policy can only list directories, search, read files, and submit a context set; it has no edit, patch, test, or repair-verification action. Its estimands are therefore File F1, gold contact, tool calls, and context tokens. Gold contact is often a prerequisite for a correct repair, but it is not sufficient: reasoning, editing, and validation may still fail after the right file is found. Conversely, the historical resolving change used to construct a positive sample is a project outcome shared by construction, not the evaluated agent's outcome, so it cannot serve as a varying patch-success label. The benchmark also has no one-to-one mapping from its workflow-derived rows to executable SWE-bench instances. We therefore do not report either patch success or an ARB--SWE-bench success correlation.

Finally, we run a lightweight proxy-validity check for the context-acquisition efficiency (CAE) metrics. This check asks whether the new diagnostics align with observed agent-side quantities, not whether they causally prove downstream patch success. Table~\ref{tab:cae-validity} summarizes two pieces of evidence. First, in a separate 40-sample seeded intervention used for PES calibration, PES@20 correlates with actual first-gold-step reduction and event savings. Second, in observational joins, the strongest hybrid retriever's BCY@8k is positively associated with agent gold-touch and final File F1. The observational associations are weaker than the within-sample PES calibration association and vary by scaffold, which is expected because BCY measures the quality of the initial ranked context, while logged agents can still explore, miss, or recover files interactively.

\begin{center}
\begin{minipage}{0.94\textwidth}
\centering
\captionsetup{hypcap=false}
\captionof{table}{Proxy-validity checks for CAE metrics on \code{code2test}, \code{comment2context}, and \code{trace2code}. PES rows use paired control/seeded Codex GPT-5.4 trajectories. BCY rows are observational joins for RRF(Qwen3-8B+RepoMap); \(\rho\) is Spearman rank correlation.}
\label{tab:cae-validity}
\begin{tabular}{llrr}
\toprule
Metric & Target quantity & n & \(\rho\) \\
\midrule
PES@20 & First-gold-step reduction & 16 & 0.674 \\
PES@20 & Event savings & 40 & 0.529 \\
BCY@8k / OpenAI strict & Any-gold touched & 287 & 0.230 \\
BCY@8k / OpenAI strict & Final File F1 & 287 & 0.237 \\
BCY@8k / Codex GPT-5.4 & Any-gold touched & 287 & 0.172 \\
BCY@8k / Codex GPT-5.4 & Final File F1 & 287 & 0.129 \\
\bottomrule
\end{tabular}
\end{minipage}
\end{center}

A tertile sanity check shows the same direction for the hybrid retriever: moving from low to high BCY raises Codex GPT-5.4 any-gold from 0.600 to 0.792 and final File F1 from 0.268 to 0.370. These results support using CAE as a diagnostic proxy, not as a scalar causal estimate of downstream success.

\subsection{Selective Retrieval with Abstention}

The positive tasks ask retrievers to return useful files, but a deployed context engine must also decide when the repository contains no useful local context for the query. We evaluate the path-aware lexical heuristic, BM25, and Jina code embeddings with a simple abstention rule: confidence is the top retrieval score, and the retriever abstains when it falls below a threshold. All results use repo-grouped five-fold cross-validation. In each round, the threshold maximizing balanced accuracy is selected on four folds and applied once to held-out repositories.

We report two protocols because the no-gold construction matters. The \emph{mixed} protocol contains all 345 positives, 50 natural no-gold issues, and 32 counterfactual wrong-repository controls. The stricter \emph{natural evidence-only} protocol excludes the 32 counterfactual controls from both calibration and evaluation, then rebuilds the repo-grouped folds over 345 positives and 50 natural no-gold issues. This separation prevents easily detected query--repository mismatches from being mistaken for calibration on plausible but externally resolved issues.

\begin{center}
\centering
\captionsetup{hypcap=false}
\captionof{table}{Repo-grouped selective retrieval with mixed and natural evidence-only no-gold protocols. Selective success@20 counts a no-gold abstention as success and an accepted positive as success only when any gold file appears in the top 20. No-abs. success is the corresponding score when every query returns files.}
\label{tab:selective-abstention}
\scriptsize
\begin{tabular}{llrrrrrr}
\toprule
Protocol & Ranker & Pos. pass & No-gold abs. & Bal. acc & Succ.@20 & No-abs. succ. & Coverage \\
\midrule
Mixed & lexical & 0.814 & 0.598 & \textbf{0.706} & \textbf{0.496} & 0.461 & 0.735 \\
Mixed & Jina-0.5B & 0.446 & \textbf{0.793} & 0.620 & 0.389 & 0.452 & 0.401 \\
Mixed & BM25 & 0.672 & 0.451 & 0.562 & 0.377 & 0.429 & 0.649 \\
\midrule
Natural only & lexical & 0.423 & 0.580 & 0.502 & 0.294 & \textbf{0.499} & 0.423 \\
Natural only & Jina-0.5B & 0.377 & 0.940 & \textbf{0.658} & \textbf{0.334} & 0.489 & 0.337 \\
Natural only & BM25 & 0.188 & \textbf{0.980} & 0.584 & 0.220 & 0.463 & 0.167 \\
\bottomrule
\end{tabular}
\end{center}

The mixed protocol appears favorable to lexical abstention: selective success@20 rises from 0.461 without abstention to 0.496. That conclusion does not survive the construction-controlled analysis. Lexical abstains on all 32 counterfactual wrong-repository controls but on only 34\% of the 50 natural no-gold cases under the mixed decisions. After counterfactual controls are removed and thresholds are recalibrated within natural-only folds, no ranker improves selective success. Lexical falls from 0.499 to 0.294, Jina from 0.489 to 0.334, and BM25 from 0.463 to 0.220.

The failure has two causes. First, top-score separation on gross wrong-repository controls is much easier than distinguishing natural external resolutions from plausible local issues. Second, thresholds that maximize balanced accuracy can reject too many positives: on the natural-only protocol Jina abstains correctly on 94.0\% of no-gold cases but passes only 37.7\% of positives. Thus the selective track currently establishes a negative result: raw top retrieval scores are not sufficient repository-independent abstention signals for natural no-gold issues. The counterfactual stratum remains useful as a separate sanity check, but must not be pooled into the primary natural-abstention claim.

\subsection{Budget-Dependent Retrieval on \code{edit2ripple}}

The positive-only \code{edit2ripple} expansion is evaluated with the standard retrieval protocol rather than abstention. Figure~\ref{fig:edit2ripple-bcy} visualizes canonical file-level BCY on 58 evidence-backed samples; Appendix Table~\ref{tab:edit2ripple-full} reports the rank and budget metrics. Qwen3-4B has the best broad file coverage (Recall@20 0.7112). Among the non-provisional runs, Qwen3-4B leads at the fully packed 4k budget and Qwen3-8B leads at the fully packed 8k budget. This Qwen ordering reversal is the robust budget-dependent result. The recorded pplx run is slightly higher at 4k but remains provisional because of its tokenizer warning. The 16k and 32k values become top-20-prefix lower bounds whenever the stored files do not fill the budget, so their apparent winners are reported descriptively rather than as definitive comparisons.

\begin{center}
\centering
\captionsetup{hypcap=false}
\begin{tikzpicture}[x=3.7cm,y=7.0cm,font=\small]
  \foreach \y/\label in {0.2/0.2,0.3/0.3,0.4/0.4,0.5/0.5,0.6/0.6,0.7/0.7} {
    \draw[gray!18] (0,\y) -- (3,\y);
    \node[anchor=e,font=\scriptsize] at (-0.06,\y) {\label};
  }
  \draw[->] (0,0.1) -- (3.2,0.1);
  \draw[->] (0,0.1) -- (0,0.76);
  \foreach \x/\label in {0/4k,1/8k,2/16k,3/32k} {
    \draw (\x,0.095) -- (\x,0.105);
    \node[anchor=north,font=\scriptsize] at (\x,0.085) {\label};
  }
  \node[anchor=north] at (1.5,0.035) {Context budget (tokens)};
  \node[rotate=90] at (-0.42,0.43) {BCY};

  \draw[gray!55,line width=0.55pt] plot coordinates {(0,0.2773) (1,0.3851) (2,0.4928) (3,0.6408)};
  \draw[gray!55,line width=0.55pt] plot coordinates {(0,0.2917) (1,0.3980) (2,0.5244) (3,0.5690)};
  \draw[gray!55,line width=0.55pt] plot coordinates {(0,0.3218) (1,0.4468) (2,0.5187) (3,0.5876)};
  \draw[gray!55,line width=0.55pt] plot coordinates {(0,0.2040) (1,0.3592) (2,0.5187) (3,0.5848)};
  \draw[gray!55,line width=0.55pt] plot coordinates {(0,0.1221) (1,0.2428) (2,0.3103) (3,0.4641)};

  \draw[orange!90!black,line width=1.5pt] plot coordinates {(0,0.3707) (1,0.5043) (2,0.6078) (3,0.7112)};
  \draw[blue!75!black,dashed,line width=1.5pt] plot coordinates {(0,0.3333) (1,0.5129) (2,0.5963) (3,0.6767)};
  \draw[green!55!black,dash dot,line width=1.5pt] plot coordinates {(0,0.3865) (1,0.4842) (2,0.6078) (3,0.6652)};
  \foreach \x/\y in {0/0.3707,1/0.5043,2/0.6078,3/0.7112} \fill[orange!90!black] (\x,\y) circle (1.5pt);
  \foreach \x/\y in {0/0.3333,1/0.5129,2/0.5963,3/0.6767} \fill[blue!75!black] (\x,\y) circle (1.5pt);
  \foreach \x/\y in {0/0.3865,1/0.4842,2/0.6078,3/0.6652} \fill[green!55!black] (\x,\y) circle (1.5pt);

  \draw[orange!90!black,line width=1.5pt] (0.05,0.805) -- (0.25,0.805);
  \node[anchor=west,font=\scriptsize] at (0.28,0.805) {Qwen3-4B};
  \draw[blue!75!black,dashed,line width=1.5pt] (1.05,0.805) -- (1.25,0.805);
  \node[anchor=west,font=\scriptsize] at (1.28,0.805) {Qwen3-8B};
  \draw[green!55!black,dash dot,line width=1.5pt] (2.05,0.805) -- (2.25,0.805);
  \node[anchor=west,font=\scriptsize] at (2.28,0.805) {pplx-4B$^{\dagger}$};

  \node[above=2pt,font=\scriptsize,green!45!black] at (0,0.3865) {pplx$^{\dagger}$};
  \node[above=2pt,font=\scriptsize,blue!70!black] at (1,0.5129) {Qwen3-8B};
  \node[above=2pt,font=\scriptsize] at (2,0.6078) {tie};
  \node[above=2pt,font=\scriptsize,orange!85!black] at (3,0.7112) {top-20 LB};
\end{tikzpicture}

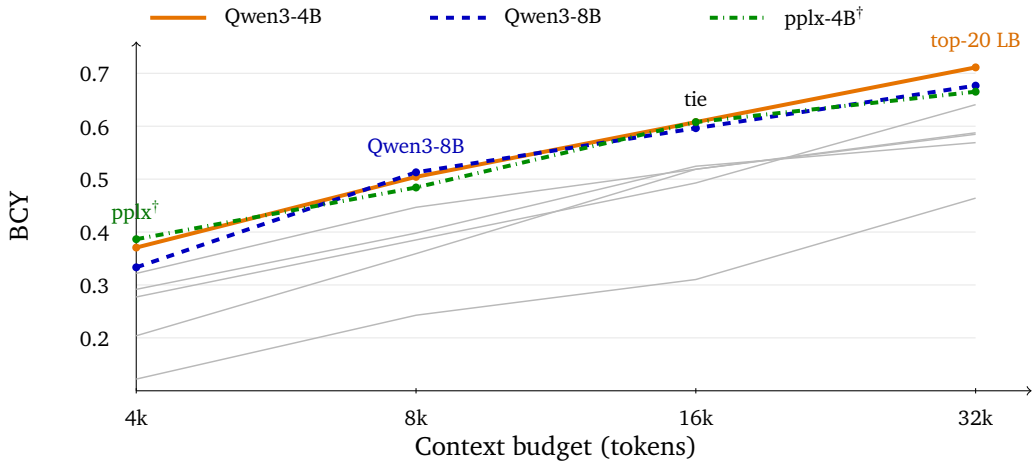
\captionof{figure}{Canonical BCY curves on \code{edit2ripple}. Gray lines show the other five baselines. The verified Qwen ordering reverses between fully packed 4k and 8k budgets. $\dagger$ marks the provisional pplx run; 16k/32k points can be lower bounds from stored top-20 rankings.}
\label{fig:edit2ripple-bcy}
\end{center}

The figure also shows why \code{edit2ripple} is not just an easier lexical variant. Lexical matching is competitive with RepoMap by MRR and BCY, but neither dominates the embedding models on Recall@20. RepoMap improves over BM25 and helps identify ripple-effect files through repository structure, yet remains behind the strongest embedding models at the fully packed 4k and 8k operating points. A single BCY point would hide the verified reversal between Qwen3-4B and Qwen3-8B. We retain the provisional pplx curve and the 32k lower bounds for transparency, but do not use either to support a definitive winner claim.

\subsection{\code{trace2code} Reverses the Aggregate Picture}

\code{trace2code} is the strongest evidence that semantic embeddings alone are insufficient. RepoMap leads this task by MRR and Recall@20. Lexical retrieval is also stronger than most embedding baselines on MRR. Qwen3-4B, despite leading overall MRR, has weak \code{trace2code} MRR (0.0827).

\begin{center}
\centering
\captionsetup{hypcap=false}
\captionof{table}{\code{trace2code} task-level results.}
\label{tab:trace2code}
\begin{tabular}{lrrrr}
\toprule
Model & R@5 & R@10 & R@20 & MRR \\
\midrule
aider-style-repomap & 0.4488 & 0.7195 & 0.8366 & 0.2742 \\
lexical & 0.3432 & 0.4818 & 0.6964 & 0.2075 \\
Qwen3-8B & 0.2442 & 0.5281 & 0.7970 & 0.1654 \\
BM25 & 0.2228 & 0.3218 & 0.4934 & 0.1638 \\
pplx-embed-v1-4b & 0.1601 & 0.2855 & 0.5380 & 0.1372 \\
nomic-embed-code & 0.1584 & 0.1733 & 0.3581 & 0.0871 \\
Qwen3-4B & 0.0743 & 0.1815 & 0.5050 & 0.0827 \\
jina-code-embeddings-0.5b & 0.0693 & 0.1188 & 0.2607 & 0.0607 \\
\bottomrule
\end{tabular}
\end{center}

Failure traces often expose tests and local symbols, not the implementation file that should change. Structure-aware ranking can connect these visible signals to nearby source files.

\section{Rank and Error Analysis}

Aggregate MRR measures early precision, but it does not fully describe whether a model can keep a gold file within a later reranking or context-selection depth. First-gold depth analysis shows this tradeoff.

\begin{center}
\centering
\captionsetup{hypcap=false}
\captionof{table}{First-gold rank depth. Any@k is the fraction of samples where at least one gold file appears in the top k.}
\label{tab:first-gold-depth}
\begin{tabular}{lrrrrr}
\toprule
Model & Any@5 & Any@20 & Med. Rank & Trace Any@20 & Trace Med. \\
\midrule
Qwen3-4B & 0.3589 & 0.6864 & 11 & 0.5347 & 20 \\
Qwen3-8B & 0.4077 & 0.7944 & 7 & 0.8614 & 9 \\
pplx-4B & 0.3310 & 0.6794 & 10 & 0.5941 & 18 \\
RepoMap & 0.3659 & 0.7178 & 9 & 0.9010 & 6 \\
Jina-0.5B & 0.2683 & 0.5226 & 19 & 0.2871 & 29 \\
Nomic & 0.3101 & 0.5784 & 15 & 0.3861 & 25 \\
Lexical & 0.2474 & 0.5470 & 17 & 0.7525 & 8 \\
\bottomrule
\end{tabular}
\end{center}

Qwen3-4B has the highest overall MRR, but Qwen3-8B has broader top-20 coverage (0.7944 vs. 0.6864). This suggests that the 4B model places its successful hits earlier, while the 8B model retrieves at least one gold file for more samples by depth 20.

Only 22 of the 287 \code{code2test}, \code{comment2context}, and \code{trace2code} samples are missed by every reported baseline at top 20. This supports the interpretation that \arb{} exposes complementary retrieval behavior, not just uniformly impossible examples.

\subsection{\code{trace2code} Error Patterns}

Representative \code{trace2code} examples show both structure wins and embedding wins. In two Gin samples whose gold files are \code{response\_writer.go} and \code{errors.go}, RepoMap retrieves the gold source file in the top 20 while Qwen3-4B and Qwen3-8B miss it. The failure excerpts mention tests and compile failures, so semantic models often rank visible test files or generic repository metadata above the implementation file. RepoMap can use structural proximity between tests and source files to recover the root-cause file.

The reverse also occurs. In a Caddy authentication failure, pplx-embed-v1-4b ranks the gold \code{caddyauth.go} file first while RepoMap misses it at depth 20. These examples prevent a simplistic ``RepoMap beats embeddings'' conclusion. The stronger conclusion is that semantic and structural signals are complementary.

\subsection{\code{comment2context} Error Patterns}

\code{comment2context} exposes the given-file trap. The reviewed file is already available to the agent, so returning it is not sufficient. In a Hypothesis seed-printing sample, the query points at the reviewed test while the gold context is in \code{core.py} and the conjecture engine. RepoMap ranks the given test file first and misses the gold files at depth 20.

\subsection{\code{code2test} Error Patterns}

\code{code2test} failures often require mapping implementation changes to project-specific test organization. An etcd sample includes nine implementation files and two existing test files in the query summary, but all seven baselines miss the gold e2e control-plane tests and cluster framework file at depth 20. These examples show that source-to-test retrieval is not only semantic similarity.

\section{Candidate Filter Ablation}

The official leaderboard uses \code{candidate\_filter=all\_files}. We also run a diagnostic candidate-filter ablation for vectorless baselines. Restricting candidates to \code{tests\_only} sharply improves \code{code2test}, but it collapses \code{trace2code} to zero because source-file golds are removed. Conversely, \code{code\_only} improves vectorless overall MRR, especially by reducing distracting non-source candidates. These results are useful diagnostically, but they should not replace the official \code{all\_files} leaderboard.

\section{Limitations}

The benchmark is diagnostic, not web-scale. It contains 427 samples: 345 positive retrieval examples, 50 natural no-gold examples, and 32 counterfactual wrong-repository controls. The 25 repositories cover six primary languages, but this is not a large statistical benchmark covering all languages and ecosystems. Repository frequency is uneven: Gin supplies 25.5\% of positive samples and the four largest repositories supply 58.8\%; Appendix Table~\ref{tab:repo-coverage} makes the full language and task distribution explicit. We therefore report repository-macro rank metrics alongside sample- and task-weighted results; the change in the Recall@20 leader under repository weighting is evidence that aggregate rankings should not be treated as repository-uniform.

\arb{} is primarily file-level. This deliberately isolates document reachability, but file exposure is not the same as localizing the useful region. In the positive benchmark, 51.8\% of sample--gold-file occurrences exceed 500 lines, while the median labeled span evidence on the 287 span-annotated samples occupies only 4.7\% of its file. BCY can therefore give full file-level credit even when a downstream agent still has substantial within-file search to perform. Our span sensitivity analysis partially quantifies this gap, but it is not a full span-retrieval leaderboard: span annotations currently cover only \code{code2test}, \code{comment2context}, and \code{trace2code}; the stored diagnostic uses legacy character-budget chunk rankings rather than canonical token packing; and file-only methods such as RepoMap do not expose line intervals. The benchmark also does not evaluate exact edit localization or patch generation, so it does not measure end-to-end agent success or establish that a retrieval miss causes a repair failure.

The corpus is fixed at the base commit. The main static-retrieval leaderboard does not evaluate dynamic tool use, iterative retrieval, or agent-driven exploration over multiple turns; the logged trajectory analysis is reported separately as a context-selection and cost comparison. Appendix~\ref{app:closed-tool} reports an additional closed-tool diagnostic in which Codex can access repository content only through evaluator-mediated \code{list\_dir}, \code{grep}, and \code{read\_file} actions. We keep the non-adaptive scripted-grep policy out of the main leaderboard because it is a deliberately weak tool-path lower bound rather than a realistic agent or an estimate of grep's capability. The trajectory cost numbers and seed-intervention results measure context acquisition, not a full causal attribution of downstream patch failures. The seed pilot also contains one trajectory per sample--arm pair, and its Codex CLI metadata records neither an explicit temperature nor a random seed. Consequently, the reported paired deltas include policy-sampling variance and establish neither confidence intervals nor statistically significant differences between retrieval arms.

Gold labels approximate files that need to be read and are supported by released construction evidence. Some tasks may have alternative useful files that are not labeled as gold.

The agentic-relevance relations are deterministic diagnostic annotations rather than independently validated latent classes. Their multi-label form reflects genuine overlap, while the primary projection introduces task-specific thresholds for descriptive slicing. These fixed thresholds are benchmark conventions, not learned empirical boundaries, and they do not enter the benchmark's main retrieval scores. Among the 287 taxonomy-labeled samples, causal-indirect is exactly the \code{trace2code} task, and structural-indirect primary cases occur only in \code{comment2context}; relevance-type results are therefore confounded with task composition. We publish secondary labels and the witness fields documented in Appendix~\ref{app:sample-schema}, but do not treat the primary taxonomy table as evidence that a relevance relation alone causes a retriever's performance.

\section{Future Work}

The most direct technical next step is hybrid retrieval. The reported results show that embeddings, lexical matching, and RepoMap have complementary strengths. A future system should combine semantic vectors, lexical/path signals, repository graphs, source-test relations, and task-aware reranking.

Another direction is a larger release with more languages, repositories, and workflow signals. Potential new tasks include issue-to-code retrieval, bug-report-to-root-cause retrieval, migration-to-affected-files retrieval, and API-usage-to-implementation retrieval.

A complementary direction is query rewriting. The benchmark preserves workflow-derived signals such as PR summaries, review comments, and failure excerpts. Future releases could add paired developer-facing rewrites of the same samples, allowing evaluation of whether retrievers are robust to phrasing and whether normalized coding-assistant queries change the relative value of semantic, lexical, and structural signals.

Finally, \arb{} should be extended in both directions. At finer granularity, future releases should add span annotations to \code{edit2ripple}, require every method to emit line-resolved intervals, and evaluate span yield under the same canonical token packing as BCY. At the downstream level, the clean test is a paired intervention on executable issue-resolution instances: hold the model, scaffold, tool and token budgets, and sampling configuration fixed; vary only the initial context among no seed, random non-gold, lexical, embedding, RRF, and oracle-gold arms; then score patches with the official test harness. Patch resolution rate is the primary endpoint, with search calls, tokens, and time-to-first-gold as secondary endpoints. Paired significance tests should compare each seed to no-seed and random controls. A separate failure-attribution analysis over aligned SWE-bench trajectories could then test whether missing necessary files predicts patch failure after controlling for repository and task difficulty.

\section{Conclusion}

\arb{} evaluates an upstream failure mode in coding agents: finding the repository files needed before patch generation begins, including deciding when no repository file should be returned. The results show that agentic retrieval is not solved by a single retrieval family. Qwen3-4B leads sample-weighted MRR, Qwen3-8B leads sample-weighted Recall@20, RepoMap leads aggregate BCY@8k and \code{trace2code}, and Qwen3-4B and Qwen3-8B reverse order across verified \code{edit2ripple} context budgets. Repository-macro Recall@20 changes the aggregate leader back to Qwen3-4B, underscoring that no overall ranking is repository-uniform. A simple RRF hybrid improves over the best single run on MRR and Recall@20 across \code{code2test}, \code{comment2context}, and \code{trace2code}. The natural no-gold protocol also shows that raw top retrieval scores do not yet provide useful abstention: all tested thresholds reduce selective success relative to returning a ranking for every query.

The practical implication is clear: coding-agent retrieval should be hybrid and task-aware. Semantic similarity, lexical/path matching, repository structure, and source-test relations each solve different parts of the benchmark. \arb{} provides a controlled way to measure those differences before they are hidden inside end-to-end patch generation.

\appendix
\clearpage

\section{Repository and Language Coverage}
\label{app:repository-coverage}

Table~\ref{tab:repo-coverage} enumerates every sample-bearing repository. ``Primary language'' is the repository's main implementation language; multilingual repositories are assigned one primary label for this descriptive summary. Counterfactual no-gold rows are counted under the repository whose corpus is searched, not the source repository from which the query was adapted. The distribution contains 13 Python repositories, three each for Go, Rust, and TypeScript, two Java repositories, and one JavaScript repository. Sample counts are much less balanced than repository counts: the three Go repositories contribute 142 of 345 positive samples, primarily because Gin contributes 88.

\begin{table*}[h!]
\centering
\caption{Repository, primary language, and per-task sample counts. C2T=\code{code2test}, C2C=\code{comment2context}, T2C=\code{trace2code}, E2R=\code{edit2ripple}, Nat.=natural no-gold, and Ctr.=counterfactual wrong-repository.}
\label{tab:repo-coverage}
\scriptsize
\setlength{\tabcolsep}{3.0pt}
\begin{tabular}{llrrrrrrr}
\toprule
Repository & Language & C2T & C2C & T2C & E2R & Nat. & Ctr. & All \\
\midrule
\code{HypothesisWorks/hypothesis} & Python & 0 & 1 & 0 & 0 & 0 & 5 & 6 \\
\code{astral-sh/ruff} & Rust & 0 & 6 & 0 & 0 & 1 & 3 & 10 \\
\code{caddyserver/caddy} & Go & 0 & 0 & 7 & 6 & 3 & 0 & 16 \\
\code{clap-rs/clap} & Rust & 4 & 0 & 1 & 5 & 0 & 1 & 11 \\
\code{eslint/eslint} & JavaScript & 0 & 1 & 0 & 0 & 1 & 4 & 6 \\
\code{etcd-io/etcd} & Go & 15 & 16 & 4 & 6 & 0 & 0 & 41 \\
\code{fastapi/fastapi} & Python & 3 & 0 & 0 & 0 & 0 & 0 & 3 \\
\code{gin-gonic/gin} & Go & 15 & 0 & 56 & 17 & 0 & 0 & 88 \\
\code{huggingface/diffusers} & Python & 6 & 9 & 0 & 1 & 0 & 2 & 18 \\
\code{huggingface/transformers} & Python & 15 & 13 & 0 & 8 & 0 & 0 & 36 \\
\code{ipython/ipython} & Python & 0 & 2 & 0 & 0 & 0 & 2 & 4 \\
\code{microsoft/playwright} & TypeScript & 0 & 4 & 0 & 0 & 3 & 1 & 8 \\
\code{mockito/mockito} & Java & 1 & 0 & 0 & 0 & 0 & 0 & 1 \\
\code{numpy/numpy} & Python & 0 & 1 & 0 & 0 & 5 & 1 & 7 \\
\code{pallets/click} & Python & 0 & 0 & 26 & 0 & 0 & 0 & 26 \\
\code{pydantic/pydantic} & Python & 0 & 1 & 0 & 0 & 0 & 1 & 2 \\
\code{pypa/pip} & Python & 0 & 5 & 0 & 0 & 4 & 2 & 11 \\
\code{pytest-dev/pytest} & Python & 3 & 2 & 1 & 1 & 14 & 5 & 26 \\
\code{python/mypy} & Python & 0 & 1 & 0 & 0 & 0 & 0 & 1 \\
\code{scrapy/scrapy} & Python & 0 & 3 & 0 & 0 & 3 & 0 & 6 \\
\code{spring-projects/spring-boot} & Java & 5 & 0 & 0 & 3 & 0 & 0 & 8 \\
\code{tokio-rs/tokio} & Rust & 15 & 11 & 6 & 6 & 2 & 0 & 40 \\
\code{tox-dev/tox} & Python & 0 & 2 & 0 & 0 & 0 & 3 & 5 \\
\code{vitejs/vite} & TypeScript & 19 & 1 & 0 & 5 & 14 & 2 & 41 \\
\code{vuejs/core} & TypeScript & 5 & 1 & 0 & 0 & 0 & 0 & 6 \\
\midrule
Total & -- & 106 & 80 & 101 & 58 & 50 & 32 & 427 \\
\bottomrule
\end{tabular}
\end{table*}

\section{Released Sample Schema and Examples}
\label{app:sample-schema}

The release stores one JSON object per line. The core fields below are common
across tasks; \code{query}, \code{gold}, and evidence fields are task-specific.
For readability, the examples are YAML-style projections of real released
records. We show the full query and all fields that determine the target or
justify abstention, while omitting housekeeping metadata such as generation
timestamps and confidence flags. Line wrapping is typographic only. The
placeholder for a reproduction-log path removes a non-semantic artifact
location, not a query or relevance witness.

\begingroup
\footnotesize
\begin{verbatim}
{
  "id": string,
  "task_type": "code2test" | "comment2context" | "trace2code"
               | "edit2ripple" | "abstention",
  "repo": "owner/name",
  "base_commit": git_sha,
  "candidate_corpus"?: {
    "type": "repo_at_base_commit",
    "base_commit": git_sha
  },
  "query": task_specific_object,
  "gold": task_specific_object,
  "gold_blocks"?: [{
    "path": path, "start_line": int, "end_line": int,
    "kind": string, "symbol": string
  }],
  "gold_spans"?: [{
    "path": path, "start_line": int, "end_line": int
  }],
  "metadata": {
    "evidence"?: object,
    "gold_evidence"?: [object],
    "evidence_summary"?: string,
    "evidence_urls"?: [url]
  }
}
\end{verbatim}
\endgroup

\begin{table*}[h!]
\centering
\caption{Task-specific target and evidence fields in the released JSONL. Optional
span fields provide finer-grained supporting regions but do not change the
file-level gold set.}
\label{tab:sample-witness-fields}
\scriptsize
\setlength{\tabcolsep}{4pt}
\begin{tabularx}{\linewidth}{l>{\raggedright\arraybackslash}p{0.27\linewidth}X}
\toprule
Task & Target fields & Relevance or abstention witnesses \\
\midrule
\code{code2test} &
\code{gold.related\_tests} &
\code{metadata.evidence.source}, \code{signals},
\code{implementation\_files}, and \code{related\_tests} \\
\code{comment2context} &
\code{gold.given\_files};
\code{gold.must\_context\_files[].path} &
\code{must\_context\_files[].evidence};
\code{metadata.evidence.gold\_definition}, review-comment and response-commit
identifiers, and \code{post\_comment\_commits} \\
\code{trace2code} &
\code{gold.root\_cause\_files} &
\code{metadata.evidence.signals}, \code{run\_status},
\code{test\_files}, and \code{combined\_log} \\
\code{edit2ripple} &
\code{gold.given\_files}; \code{gold.files} &
\code{metadata.gold\_evidence[].path}, \code{signals},
\code{is\_test}, \code{additions}, and \code{deletions} \\
Natural no-gold &
\code{gold.files=[]}; \code{gold.no\_gold=true};
\code{gold.reason} &
\code{metadata.organic=true}, \code{evidence\_summary}, and
\code{evidence\_urls} \\
Counterfactual no-gold &
\code{gold.files=[]}; \code{gold.no\_gold=true};
\code{gold.reason} &
\code{metadata.organic=false}, \code{source\_repo},
\code{source\_sample\_id}, \code{pairing\_profile}, and
\code{evidence\_urls} \\
\bottomrule
\end{tabularx}
\end{table*}

\paragraph{\code{code2test}: implementation change to existing tests.}
This sample comes from \url{https://github.com/clap-rs/clap/pull/6319}.
The implementation path is visible in the query; the two test paths are not.
\begingroup
\footnotesize
\begin{verbatim}
id: f39b3eb29d5ba2f6c18b0555
task_type: code2test
repo: clap-rs/clap
base_commit: 87f57fa1db720b1f5e6b76cd5aa56c7107a3f946
query:
  pr_title: "fix(man)!: Put env support behind the env feature"
  pr_body: ""
  implementation_file_count: 1
  implementation_files: [clap_mangen/src/render.rs]
  changed_file_summary:
    "1 implementation files and 2 existing test files changed
     within 4 total files."
gold:
  fix_commit: 92e7e730743744415e44e7518ff7fb2d039c0bc6
  related_tests:
    - clap_mangen/tests/testsuite/common.rs
    - clap_mangen/tests/testsuite/roff.rs
  root_cause_files: [clap_mangen/src/render.rs]
  root_cause_symbols: []
  supporting_files: [clap_mangen/Cargo.toml]
  negative_distractors: []
metadata.evidence:
  source: pr_level_changed_implementation_and_existing_tests
  implementation_files: [clap_mangen/src/render.rs]
  related_tests:
    - clap_mangen/tests/testsuite/common.rs
    - clap_mangen/tests/testsuite/roff.rs
  signals:
    - same_pr_changed_implementation_and_tests
    - behavior_or_bug_signal
    - source_test_path_overlap
    - cross_directory_test
\end{verbatim}
\endgroup

\paragraph{\code{comment2context}: reviewed test to additional implementation context.}
This sample comes from \url{https://github.com/tokio-rs/tokio/pull/7686}.
The reviewed test is given context and is explicitly excluded from the main
gold.
\begingroup
\footnotesize
\begin{verbatim}
id: 35154a761b37d4f5767d1cfa
task_type: comment2context
repo: tokio-rs/tokio
base_commit: 2137f7d953df2c65ad254aba41e7e403905dda91
query:
  pr_title:
    "sync: return `TryRecvError::Disconnected` from
     `Receiver::try_recv` after `Receiver::close`"
  review_comment: "Please also test the `drop(tx)` case."
  given_file: tokio/tests/sync_mpsc.rs
  path: tokio/tests/sync_mpsc.rs
  line: 1
  diff_hunk_context: ""
gold:
  fix_commit: d060401f6c7dca4a20674e3ad63ad7f1b228aa31
  given_files: [tokio/tests/sync_mpsc.rs]
  must_context_files:
    - path: tokio/src/sync/mpsc/chan.rs
      evidence:
        - modified_after_review_comment
        - implementation_context_for_reviewed_change
        - explicit_behavior_or_api_dependency
  root_cause_files: [tokio/src/sync/mpsc/chan.rs]
  related_tests: [tokio/tests/sync_mpsc.rs]
  supporting_files: []
  root_cause_symbols: []
  negative_distractors:
    - tokio/src/sync/mpsc/chan.rs
    - tokio/src/sync/mpsc/list.rs
metadata.evidence:
  gold_definition: must_context_files_excluding_given_file
  review_comment_id: 2432768603
  comment_created_at: "2025-10-15T14:22:23Z"
  comment_commit_id: 8212c428d09f125e2d2c42583bee374beb5dc14d
  response_commit: fa5b8bf0b8686fa3e1afbcf68dd9cb1dce19e875
  post_comment_commits:
    - fa5b8bf0b8686fa3e1afbcf68dd9cb1dce19e875
    - e4347422b990294fbfe7e3e562121094c7eec776
    - 8212c428d09f125e2d2c42583bee374beb5dc14d
\end{verbatim}
\endgroup

\paragraph{\code{trace2code}: reproduced test failure to root-cause source.}
This sample comes from \url{https://github.com/gin-gonic/gin/pull/4404}.
The failure exposes \code{debug\_test.go}; the gold is the implementation file
\code{debug.go}.
\begingroup
\footnotesize
\begin{verbatim}
id: 0f88458fc1fe4acce078ce3f
task_type: trace2code
repo: gin-gonic/gin
base_commit: 2a794cd0b0faa7d829291375b27a3467ea972b0d
query:
  command: "go test ./."
  run_strategy: go_test_package
  source_type: local_test_reproduction
  failure_excerpt: |
    $ go test ./.
    FAIL  github.com/gin-gonic/gin [build failed]
    FAIL

    # github.com/gin-gonic/gin [github.com/gin-gonic/gin.test]
    ./debug_test.go:118:2: undefined: runtimeVersion
gold:
  fix_commit: 19b877fa50cbbb9282763099fb177a1e5cc5c850
  root_cause_files: [debug.go]
  related_tests: [debug_test.go]
  root_cause_symbols: []
  supporting_files: []
  negative_distractors: []
metadata.evidence:
  combined_log: <released reproduction log>
  failure_trace_reclassified: false
  run_status: failed_expected
  test_files: [debug_test.go]
  signals:
    - local_test_reproduction
    - test_only_patch_applied
    - failure_observed
    - compile_error
\end{verbatim}
\endgroup

\paragraph{\code{edit2ripple}: anchored edit to additional affected files.}
This sample comes from \url{https://github.com/clap-rs/clap/pull/6129}.
The anchor is given context, while the API declaration and its tests are gold.
\begingroup
\footnotesize
\begin{verbatim}
id: 8b6ab1ba0883ff8104d28524
task_type: edit2ripple
repo: clap-rs/clap
base_commit: e82e1edf76bcbddf5fe53428d297520d76a6a300
query:
  anchor_file: clap_builder/src/parser/parser.rs
  intent: |
    feat: Add default_values_if and default_values_ifs

    Fixes #5698

    feat: Add default_values_if and default_values_ifs
    to match default_values
  anchor_diff: |
    @@ -1462,7 +1462,8 @@ impl<'cmd> Parser<'cmd> {

                         if add {
                         if let Some(default) = default {
    -                        let arg_values = vec![default.to_os_string()];
    +                        let arg_values =
    +                            default.iter().map(|os_str| os_str.to_os_string()).collect();
                             let trailing_idx = None;
                             let _ = ok!(self.react(
                                 None,
gold:
  given_files: [clap_builder/src/parser/parser.rs]
  files:
    - clap_builder/src/builder/arg.rs
    - tests/builder/default_vals.rs
metadata.gold_evidence:
  - path: clap_builder/src/builder/arg.rs
    is_test: false
    additions: 70
    deletions: 2
    signals: [path_token_overlap, shared_changed_symbol]
  - path: tests/builder/default_vals.rs
    is_test: true
    additions: 331
    deletions: 0
    signals:
      - path_token_overlap
      - shared_changed_symbol
      - optional_test_ripple
\end{verbatim}
\endgroup

\paragraph{Natural no-gold: plausible local query with an external resolution.}
This sample comes from
\url{https://github.com/pytest-dev/pytest/issues/13024}. It is not a
wrong-repository control: the issue genuinely belongs to pytest, but the
query points to an upstream CPython defect and the released evidence indicates
that no pytest repository file should be returned.
\begingroup
\footnotesize
\begin{verbatim}
id: abstention_candidate__organic_issue__47097e7d8f29ad1aa2e839b5
task_type: abstention
repo: pytest-dev/pytest
base_commit: a0e3a49f194175bec96a551791b2b02b23e06a09
query:
  source: issue
  text: |
    Issue title: SystemError occured in Windows Python 3.13.0
    Issue body excerpt: I encountered the following error when using
    Pytest 8.3.3 in Windows Python 3.13.0 in GitHub Actions:

    ```
    > Run poetry run pytest --cov-append --cov-report=xml
      --junitxml=junit.xml
    Exception ignored on threading shutdown:
    Traceback (most recent call last):
      File "C:\hostedtoolcache\windows\Python\3.13.0\x64\
      Lib\threading.py", line 1524, in _shutdown
        if _main_thread._handle.is_done()
          and _is_main_interpreter():
    SystemError: <method 'is_done' of '_thread._ThreadHandle'
      objects> returned a result with an exception set
    Error: Process completed with exit code 1.
    ```

    I saw a similar issue in the cpython repository:
    https://github.com/python/cpython/issues/125842.
    I wonder if this is just a bug with Python 3.13.0 and not pytest's.
gold:
  files: []
  no_gold: true
  reason: upstream_dependency
metadata:
  organic: true
  evidence_summary:
    "Maintainer resolution evidence classifies the issue as no
     repository-local fix; automatic checks found no issue closing
     commit event."
  evidence_urls:
    - https://github.com/pytest-dev/pytest/issues/13024
\end{verbatim}
\endgroup

\section{Embedding Baseline Implementation}
\label{app:embedding-implementation}

We load each checkpoint through SentenceTransformers and call \code{encode} with output normalization enabled. No run supplies \code{query\_prefix}, \code{passage\_prefix}, or a SentenceTransformers \code{prompt\_name}. Some checkpoints ship recommended retrieval prompts, but those templates are not selected automatically because their \code{default\_prompt\_name} is unset. The reported rows consequently measure a common prompt-free interface. This choice improves protocol uniformity but should not be interpreted as each model's best task-specific configuration.

\begin{table}[h!]
\centering
\caption{Embedding baseline implementation. Revision gives the first 12 hexadecimal characters of the Hugging Face commit. All rows use no added prompt, L2-normalized output, right truncation, and max chunk-to-file aggregation.}
\label{tab:embedding-implementation}
\small
\begin{tabular}{llllrr}
\toprule
Checkpoint & Revision & Pooling & Similarity & Dim. & Max tokens \\
\midrule
Qwen3-Embedding-4B & \code{5cf2132abc99} & last token & cosine & 2560 & 40,960 \\
Qwen3-Embedding-8B & \code{1d8ad4ca9b3d} & last token & cosine & 4096 & 40,960 \\
jina-code-embeddings-0.5b & \code{4db235132daf} & last token & cosine & 896 & 32,768 \\
pplx-embed-v1-4b & \code{2cd0f789519b} & mean & cosine & 2560 & 32,768 \\
nomic-embed-code & \code{11114029805c} & last token & cosine & 3584 & 32,768 \\
\bottomrule
\end{tabular}
\end{table}

Each corpus chunk is rendered as follows; the \code{symbol} line is omitted when no symbol is available.
\begin{verbatim}
path: {path}
kind: {kind}
symbol: {symbol}
content:
{text}
\end{verbatim}
Queries use the benchmark's evaluation query text without an added instruction. All tokenizers use right padding and right truncation. Table~\ref{tab:embedding-implementation} reports the effective SentenceTransformers sequence cap, i.e., the active limit after reconciling the checkpoint's model and tokenizer settings. Chunks are scored by cosine similarity. File-level rankings use max aggregation over chunks: for query $q$, file $f$, and its chunks $C_f$,
\[
s(q,f)=\max_{c\in C_f}\left\langle
\frac{e(q)}{\left\|e(q)\right\|_2},
\frac{e(c)}{\left\|e(c)\right\|_2}
\right\rangle.
\]
Ties are resolved deterministically by path and chunk identifier, after which the evaluator retains the first occurrence of each file.

The reported \code{edit2ripple} runs use an NVIDIA H20 GPU with embedding and query batch sizes of 8. Exact checkpoint identifiers and software versions are retained in the released run metadata; runtime is not compared as a model-quality result because it depends on hardware and cache state.

One implementation caveat applies to pplx-embed-v1-4b. Transformers 4.57.6 emitted a tokenizer warning recommending \code{fix\_mistral\_regex=true}; the current evaluator did not forward that tokenizer keyword. The reported pplx row therefore corresponds to the tokenizer as loaded in the recorded run and is provisional pending a corrected-tokenizer rerun. We do not claim that the regex fix was applied. The other four checkpoints did not emit this warning.

\section{Full \code{edit2ripple} Results}

Table~\ref{tab:edit2ripple-full} provides the numerical values behind Figure~\ref{fig:edit2ripple-bcy}. BCY uses the canonical token-based file-packing protocol at each budget and the stored top-20 file rankings. The 4k and 8k budgets are fully packed in the recorded runs; 16k and 32k values are lower bounds whenever the stored prefix underfills the budget. The pplx row remains provisional pending the corrected-tokenizer rerun described above.

\begin{table*}[h]
\centering
\caption{\code{edit2ripple} retrieval results on 58 evidence-backed samples. $\dagger$ marks the provisional pplx run; BCY@32k is a top-20-prefix lower bound (LB).}
\label{tab:edit2ripple-full}
\scriptsize
\setlength{\tabcolsep}{5.8pt}
\begin{tabular}{llrrrrrrr}
\toprule
Model & Family & R@5 & R@20 & MRR & BCY@4k & BCY@8k & BCY@16k LB & BCY@32k LB \\
\midrule
Qwen3-Embedding-4B & embedding & 0.4813 & \textbf{0.7112} & 0.2791 & 0.3707 & 0.5043 & \textbf{0.6078} & \textbf{0.7112} \\
Qwen3-Embedding-8B & embedding & \textbf{0.4914} & 0.6825 & 0.2652 & 0.3333 & \textbf{0.5129} & 0.5963 & 0.6767 \\
pplx-embed-v1-4b$^{\dagger}$ & embedding & 0.4670 & 0.6782 & \textbf{0.2877} & \textbf{0.3865} & 0.4842 & \textbf{0.6078} & 0.6652 \\
jina-code-embeddings-0.5b & embedding & 0.3678 & 0.6466 & 0.2414 & 0.2773 & 0.3851 & 0.4928 & 0.6408 \\
nomic-embed-code & embedding & 0.3807 & 0.5733 & 0.2860 & 0.2917 & 0.3980 & 0.5244 & 0.5690 \\
\midrule
lexical & lexical & 0.4124 & 0.5876 & 0.2428 & 0.3218 & 0.4468 & 0.5187 & 0.5876 \\
RepoMap & structure & 0.3635 & 0.6164 & 0.2344 & 0.2040 & 0.3592 & 0.5187 & 0.5848 \\
BM25 & lexical & 0.2069 & 0.4986 & 0.1536 & 0.1221 & 0.2428 & 0.3103 & 0.4641 \\
\bottomrule
\end{tabular}
\end{table*}

\section{Closed-Tool Context Acquisition Diagnostic}
\label{app:closed-tool}

The main leaderboard evaluates static file ranking. To test whether \arb{} can also support an agent-native context-acquisition protocol, we ran a closed-tool diagnostic in which the policy must iteratively request repository information. The evaluated policy is isolated from the corpus filesystem: Codex runs inside a Docker container whose workspace contains only the current prompt, action schema, and prior observations. The benchmark corpus is not mounted into the container. At each turn, Codex returns a structured JSON action from \code{list\_dir}, \code{grep}, \code{read\_file}, and \code{submit}; the evaluator executes the action outside the container and returns only the resulting observation. We audit Codex JSON events for shell or unsupported tool use and count such events as protocol violations.

Table~\ref{tab:closed-tool-overall} reports the full 287-sample three-task diagnostic. The deterministic scripted-grep row is a non-adaptive lower bound and tool-path smoke test: it uses the same evaluator-mediated budgets, but it is not an estimate of what grep can achieve under a competent policy. Before observing any result, the script selects at most eight high-weight standalone query tokens or quoted phrases, runs independent substring searches, scores files using match counts, ranks, and path-token bonuses, reads the highest-scoring candidates, and submits at most three files. It never composes Boolean queries, follows symbols, maps source files to tests, or rewrites a pattern from an observation. The Docker-isolated Codex row is the adaptive policy result.

\begin{table}[h]
\centering
\caption{Closed-tool context-acquisition diagnostic on the 287 \code{code2test}, \code{comment2context}, and \code{trace2code} samples. Both rows use evaluator-mediated repository access; scripted grep is a non-adaptive lower bound, not a main agent baseline.}
\label{tab:closed-tool-overall}
\small
\begin{tabular}{lrrrrrrr}
\toprule
Policy & n & Final F1 & Final R & Final P & Any-gold & Calls & Viol. \\
\midrule
scripted grep & 287 & 0.0719 & 0.1144 & 0.0557 & 0.2683 & 15.79 & 0 \\
Codex closed-tool & 287 & 0.3279 & 0.4692 & 0.2712 & 0.5470 & 3.82 & 0 \\
\bottomrule
\end{tabular}
\end{table}

The scripted row's 0.0719 Final F1 should not be read as evidence that the benchmark categorically suppresses lexical retrieval. Final F1 scores only the three submitted files and is low for static methods generally: on the same samples, matched top-3 File F1 is 0.0721 for the path-aware Lexical baseline, 0.0841 for Grep strict, 0.1102 for RepoMap, and 0.1247 for Qwen3-4B (Table~\ref{tab:context-selection}). At a ranking depth of 20, however, Lexical reaches 0.4750 gold-file Recall and 0.5470 Any@20. Its Recall@20 also varies substantially by workflow: 0.2469 on \code{code2test}, 0.4979 on \code{comment2context}, and 0.6964 on \code{trace2code}. Thus the benchmark is intentionally difficult for literal query-to-file matching when relevance is indirect, but it does not uniformly disadvantage lexical signals. The low scripted score primarily demonstrates that a fixed sequence of uncomposed substring searches is a poor context-selection policy; it is not a fairness test for the benchmark or an upper bound on interactive lexical search.

Table~\ref{tab:closed-tool-task} breaks down the Docker-isolated Codex row by task. The result is diagnostic rather than a replacement for the static leaderboard. It shows that a controlled interactive policy can recover substantially more context on \code{trace2code}, where failure output provides strong localization clues, while \code{comment2context} remains difficult even with adaptive tool use.

\begin{table}[h]
\centering
\caption{Docker-isolated Codex closed-tool results by task.}
\label{tab:closed-tool-task}
\small
\begin{tabular}{lrrrrrr}
\toprule
Task & n & Final F1 & Final R & Final P & Any-gold & Read toks \\
\midrule
\code{code2test} & 106 & 0.2387 & 0.3601 & 0.1887 & 0.4151 & 2391.3 \\
\code{comment2context} & 80 & 0.1325 & 0.1583 & 0.1292 & 0.2625 & 2026.2 \\
\code{trace2code} & 101 & 0.5762 & 0.8300 & 0.4703 & 0.9109 & 2477.6 \\
\bottomrule
\end{tabular}
\end{table}

We also report a post-hoc tool-budget prefix diagnostic in Table~\ref{tab:closed-tool-budget}. For each completed Codex trajectory, we score the files read within the first \(B\) evaluator-mediated tool calls as the acquired context. This is not a counterfactual rerun under a smaller budget; it measures how quickly the recorded policy acquired useful files. The main gain occurs between two and four tool calls, after which the prefix score saturates because most runs have already submitted or exhausted useful reads.

\begin{table}[h]
\centering
\caption{Post-hoc budget prefix diagnostic for Docker-isolated Codex trajectories on \code{code2test}, \code{comment2context}, and \code{trace2code}. Prefix File F1 scores files read within the first \(B\) tool calls.}
\label{tab:closed-tool-budget}
\small
\begin{tabular}{rrrrrrr}
\toprule
Budget & n & Prefix F1 & Prefix R & Prefix P & Any-gold & Read toks \\
\midrule
2 & 287 & 0.1863 & 0.2340 & 0.1707 & 0.2578 & 1409.0 \\
4 & 287 & 0.3276 & 0.4692 & 0.2709 & 0.5470 & 2319.9 \\
6 & 287 & 0.3276 & 0.4692 & 0.2709 & 0.5470 & 2319.9 \\
8 & 287 & 0.3276 & 0.4692 & 0.2709 & 0.5470 & 2319.9 \\
\bottomrule
\end{tabular}
\end{table}

This diagnostic supports two limited claims. First, \arb{} can evaluate more than single-shot static ranking: the same gold files and trajectory metrics can score an evaluator-mediated interactive retrieval policy. Second, the large task spread confirms that closed-tool context acquisition is not homogeneous. The diagnostic does not claim that deterministic grep is a meaningful agent baseline, nor that the closed-tool protocol captures the full behavior of production coding agents with editing, testing, and long-lived memory.

\renewcommand{\bibfont}{\footnotesize}
\bibliographystyle{plainnat}
\bibliography{references}

\end{document}